\documentclass[page-classic]{epl2} % for one column style

\usepackage{amsmath}
\usepackage{amssymb}
\usepackage{bm}
\usepackage{graphicx}
\usepackage{subfigure}

\usepackage{color}
\usepackage{hyperref}

\title{
  Vulnerability and Resilience of Social Engagement:\\
  Equilibrium Theory
}

\author{Shang-Nan Wang\inst{1,3} \and Luan Cheng\inst{2} \and Hai-Jun Zhou\inst{1,3,4}}

\institute{
  \inst{1}
  CAS Key Laboratory for Theoretical Physics, Institute of Theoretical Physics, Chinese Academy of Sciences, Beijing 100190, China \\
  \inst{2}
  Institute of Theoretical Physics, Dalian University of Technology, Dalian 116024, China \\
  \inst{3}
  School of Physical Sciences, University of Chinese Academy of Sciences, Beijing 100049, China \\
  \inst{4}
  Synergetic Innovation Center for Quantum Effects and Applications, Hunan Normal University, Changsha 410081, China
}


\abstract{
  Social networks of engagement sometimes dramatically collapse. A widely adopted paradigm to understand this catastrophe dynamics is the threshold model but previous work only considered the irreversible $K$-core pruning process and the resulting kinetic activity patterns. Here we study the network alliance problem as a simplified model of social engagement by equilibrium statistical mechanics. Our theory reveals that the surviving kinetic alliances are out-of-equilibrium and atypical configurations which may become highly vulnerable to single-node-triggered cascading failures as they relax towards equilibrium.  Our theory predicts that if the fraction of active nodes is beyond a certain critical value, the equilibrium (typical) alliance configurations could be protected from cascading failures by a simple least-effort local intervention strategy. We confirm these results by extensive Monte Carlo simulations.
}

\begin{document}

%\date{\today}

\maketitle

The proper functioning of online and offline social networks requires the active engagement of its members. But social engagement is largely a collective phenomenon as individual agents influence and are influenced by their network neighbors~\cite{Granovetter-1978,Seidman-1983}.  Small variations of environmental parameters and localized state perturbations sometimes trigger global disruptions of social engagement, such as the rapid decline of online platforms~\cite{Garcia-etal-2013}, breakdown of social trust in vaccination~\cite{OConnor-Weatherall-2019}, military mutiny and regime shift~\cite{Gallopin-2019}, and many others~\cite{Long-Gao-2019,Rand-etal-2014,Dai-etal-2012,Banani-etal-2017,Gladwell-2000,Scheffer-2009}. Understanding the collapse of social engagement and exploring effective intervention mechanisms are research issues of great practical relevance~\cite{Lehmann-Ahn-2018,Vespignani-2012,Castellano-Fortunato-Loreto-2009}.

Previous theoretical studies modeled the disruption of social engagement as an irreversible threshold dynamics~\cite{Chalupa-Leath-Reich-1979,Pittel-etal-1996,Watts-2002,Dorogovtsev-etal-2006,Farrow-etal-2007,Zhao-Zhou-Liu-2013,Shrestha-Moore-2014,Baxter-etal-2015,Yuan-etal-2016,Morone-etal-2018,Rizzo-2018,Xie-etal-2019}. Starting from an initial random pattern in which the nodes are active with probability $p$, the system goes through a damage cascading process essentially identical to $K$-core pruning, with active nodes decaying to inactive if they have too few active neighbors~\cite{Chalupa-Leath-Reich-1979,Pittel-etal-1996}. The final configurations are extremely sensitive to $p$ if it is close to a certain critical value, at which an extensive drop in the network's activity level may occur~\cite{Dorogovtsev-etal-2006,Farrow-etal-2007,Zhao-Zhou-Liu-2013}.  However, this kinetic framework neglects a crucial aspect of social engagement, namely the activity configurations are far from being random and irreversible but are the result of complicated interactions among the individual agents and are adaptive~\cite{Garcia-etal-2013,Ugander-etal-2012,DiMuro-etal-2020}.  Some large-scale empirical studies have demonstrated significant neighborhood reinforcement effects in the dynamics of health behaviors in social networks~\cite{Centola-2010,Centola-Macy-2007,Christakis-Fowler-2007,Christakis-Fowler-2008}. For example, a person who quitted smoking is very likely to resume this habit if s/he has many smoker friends.

In the present work we study social engagement as a network alliance problem by equilibrium statistical mechanics. Each node of the system can flip back and forth between the active and inactive states. We consider microscopic alliance configurations whose active nodes are supported by other active neighbors (Fig.~\ref{fig:alliance})~\cite{Kristiansen-etal-2004,Xu-etal-2018,Yeung-Saad-2013c}, and develop a mean field theory which has the novel feature of combining the dynamics of cascading propagation with equilibrium sampling of alliance configurations. We find that kinetic alliances obtained through $K$-core pruning are out-of-equilibrium atypical configurations which may become more and more unstable as they evolve towards equilibrium. The equilibrium alliance configurations can be classified into three dynamical phases depending on the abundance of active nodes [Fig.~\ref{fig:RRD6K3:a}], and we are able to predict the phase boundaries precisely for random networks. In the intermediate phase, bounded by the weak and strong tipping points, the equilibrium alliances are highly vulnerable to cascading failures but all the global collapses are suppressible by flipping a small number of inactive nodes during the cascading process.

Our theoretical and simulation results suggest that equilibrium social engagement in networks may be intrinsically fragile but also be adaptable and resilient to disturbances, which may partly explain why the exact occurrence of a collapse event is quite unpredictable. Our mean-field theory may also be useful for understanding some other collective phenomena of complex systems, such as jamming in granular matters.

\section{Theory}

\begin{figure}[t]
  \centering
  \includegraphics[angle=270,width=0.5\linewidth]{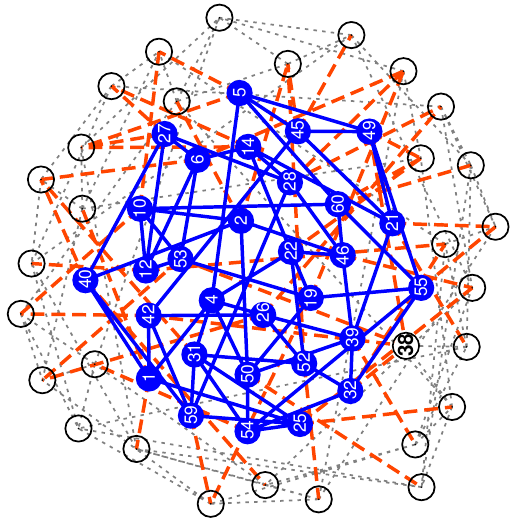}
  \caption{
    \label{fig:alliance}
    An example alliance configuration with equal number of active and inactive nodes (filled and open circles, respectively) and therefore energy density $\rho \! = \! 0.5$, for a regular random network of size $N\! = \! 60$ and degree $D\! = \! 6$, at uniform threshold $K\! =\! 4$. The links having two, one, and zero active incident nodes are drawn respectively as solid, dashed, and dotted lines. The inactive node $38$ has four active neighbors, so it can switch to be active (it is persuadable).
  }
\end{figure}

\emph{Alliance and its collapse.} Consider a network $G$ formed by $N$ nodes and some undirected links. Nodes $i$ and $j$ are neighbors if there is a link $(i, j)$ between them, and $\partial i \equiv \{j: (i, j)\! \in\! G\}$ is the neighborhood of $i$.  At any time node $i$ either actively engages in the network (state $c_i\! =\! 1$) or is inactive ($c_i\! =\! 0$), and it may switch between these states.  The engaging benefit for $i$ increases with its number $a_i$ ($\equiv\! \sum_{j\in \partial i} c_j$) of active neighbors and when $a_i$ reaches a threshold $\theta_i$ the benefit outweighs the engaging cost~\cite{Granovetter-1978,Garcia-etal-2013,Ugander-etal-2012}.  When $a_i \! < \! \theta_i$ node $i$ is always inactive, so the network configurations $\bm{c}\! \equiv\! (c_1, c_2, \ldots, c_N)$ are those which satisfy the alliance condition $a_j\! \geq\! \theta_j$ for every active node $j$. The active nodes of $\bm{c}$ are directly or indirectly supporting each other and are collectively referred to as an alliance, $A(\bm{c}) \equiv \{j: c_j\! =\! 1\}$~\cite{Kristiansen-etal-2004}. A node $i$ with many active neighbors ($a_i \! \geq \! \theta_i$) may still be inactive in network $G$ (e.g., it may be engaging in a competing network~\cite{Garcia-etal-2013}), and we consider such a node to be persuadable because it can be flipped to $c_i\! = \! 1$ (Fig.~\ref{fig:alliance}).

We define the energy of configuration $\bm{c}$ as $E(\bm{c})\! \equiv\! \sum_{i} c_i$, which is simply the size of $A(\bm{c})$. At a given energy density (relative size) $\rho\! \equiv\! E/N$ the total number $\Omega(N \rho)$ of alliances exponentially increases with system size $N$~\cite{Xu-etal-2018}, so we define the entropy density at $\rho$ as $s(\rho) \equiv (1/N) \ln \Omega$. We pick a configuration $\bm{c}$ uniformly at random from this exponential subspace and examine its sensitivity to local perturbations. If an active node $i$ drops out of $A(\bm{c})$ a damage cascading process may be triggered, during which some initially active nodes $j$ are forced to be inactive when the alliance conditions $a_j\! \geq\! \theta_j$ are violated~\cite{Granovetter-1978,Watts-2002}.  After this process finally stops the alliance might only shrink slightly (a small avalanche) or it might be extensively damaged (a collapse)~\cite{SInote2019}. We classify an avalanche of $A(\bm{c})$ as a collapse if the final energy density $\rho^\prime$ is much smaller than the initial value $\rho$, and correspondingly we regard the triggering node $i$ as a break node of $\bm{c}$. Empirically, we find that if a collapse occurs it is usually a complete one, $\rho^\prime \! = \! 0$. Therefore we take the somewhat arbitrary criterion $\rho^\prime \! <\! 0.2 \rho$ to classify a collapse (the other tested criteria of $0.1 \rho$, $0.4 \rho$ and $0.6 \rho$ all lead to identical results). The fraction $\phi$ of break nodes is computed by checking every active node of $\bm{c}$. If $\phi$ is positive then the alliance $A(\bm{c})$ is highly vulnerable to single-node perturbations.

Various local protection mechanisms are conceivable for social networks and some of them may significantly affect the order parameter $\phi$. For example every node in the reservoir set of persuadable inactive nodes has a stabilizing effect to the alliance $A(\bm{c})$. In this theoretical work we focus on the following simple least-effort local recruitment mechanism: If an active node $j$ becomes unstable ($a_j$ falls below $\theta_j$) it flips to $c_j\! = \! 0$ only if there are no enough persuadable neighbors, otherwise some persuadable neighbors $k$ are flipped to $c_k\! = \! 1$ to restore the condition $a_j\! \geq\! \theta_j$. We can study the effect of this intervention mechanism by damage cascading analysis. If the quit of a single active node $i$ still leads to a collapse, $i$ is said to be a break node of $\bm{c}$ under this locally protected cascading process. The fraction of break nodes of $\bm{c}$ is denoted as $\psi$. Obviously $\psi \! \leq \! \phi$ for any given alliance configuration $\bm{c}$. For example, all the thirty active nodes in Fig.~\ref{fig:alliance} are break nodes in the unprotected dynamics ($\phi \! =\! 0.50$), but nodes $19, 26, 58$ are no longer break nodes in the protected dynamics which recruits node $38$ ($\psi\! =\! 0.45$).

\emph{Collapse theory.} We  develop a mean field theory to analyze the vulnerability of equilibrium alliances. Each node $i$ contributes a term to $E(\bm{c})$ and it also imposes a constraint to itself and all its neighbors. Consider a link $(i, j)$ and let us for the moment neglect the energy and constraint of node $i$, so the state $(c_i, c_j)$ is affected by the energy and constraint of node $j$ only. We denote the corresponding probability distribution as $q_{j\rightarrow i}^{c_j, c_i}$, and in addition denote by $t_{j\rightarrow i}^{1,1}$ the probability that (1) $c_i\! =\! c_j\! =\! 1$ and (2) if node $i$ now flips to $c_i\! =\! 0$ the damage cascade relayed through link $(i, j)$ will only cause a tree-formed small avalanche~\cite{Zhou-2005a}. Because a large random finite-connectivity network is locally tree-like, if node $j$ is deleted from the network its neighbors will be distantly separated and be mutually independent~\cite{Mezard-Montanari-2009,Albert-Barabasi-2002}. Assuming this Bethe-Peierls factorization property, we get the following belief-propagation (BP) equation: 
\begin{subequations}
  \label{eq:BP}
  \begin{align}
    \hspace*{-0.1cm}
    q_{j\rightarrow i}^{0, c_i} & = 
    \frac{1}{z_{j\rightarrow i}} \prod_{k\in \partial j\backslash i} 
    ( q_{k\rightarrow j}^{0,0}+q_{k\rightarrow  j}^{1,0}) \; ,
    \\
    \hspace*{-0.1cm}
    q_{j\rightarrow i}^{1,c_i} & = 
    \frac{e^{-\beta}}{z_{j\rightarrow i}} 
    \sum_{\bm{c}_{\partial j\backslash i}}
    \Theta\bigl( \sum\limits_{k\in\partial j\backslash i} c_k\! +\! c_i  
    \! -\! \theta_j \bigr) \prod\limits_{k\in \partial j\backslash i}
    q_{k\rightarrow j}^{c_k,1} \; ,
  \end{align}
\end{subequations}
where $z_{j\rightarrow i}$ is the normalization constant; set $\partial j\backslash i$ contains all the neighbors of node $j$ except for $i$ and $\bm{c}_{\partial j\backslash i} \!\equiv\! \{c_k \!:\!  k \!\in\! \partial j\backslash i\}$ denotes a composite state of all the nodes in $\partial j\backslash i$; $\beta$ is the inverse temperature parameter adopted to control the energy density $\rho$; $\Theta(x) \! = \! 0$ if $x<0$ and $=\! 1$ if $x\geq 0$~\cite{Xu-etal-2018}. Similarly the probability $u_{j\rightarrow i}^{0,1}$ of node $j$ being unpersuadable ($c_j\! = \! 0$ and $a_j\! < \! \theta_j$) and $c_i\! = \! 1$ is
\begin{equation}
  u_{j\rightarrow i}^{0,1} =
  \frac{1}{z_{j\rightarrow i}} \sum\limits_{\bm{c}_{\partial j\backslash i}}
  \Theta\bigl( \theta_j \! -\! 2\! -\!
  \sum\limits_{k\in \partial j\backslash i} c_k \bigr)
  \prod_{k\in \partial j\backslash i} q_{k\rightarrow j}^{c_k, 0}
  \; .
  \label{eq:aji01}
\end{equation}

For damage cascading without local recruitments, we derive the self-consistent expression for $t_{j\rightarrow i}^{1, 1}$ by noticing that (1) the damage of node $i$ will not propagate to $j$ if initially $a_j \! > \! \theta_j$ and (2) node $j$ will flip to $c_j\! = \! 0$ if $a_j\! = \! \theta_j$ and this may then induce further damages to the alliance:
\begin{eqnarray}
  t_{j\rightarrow i}^{1,1} & = &
  \frac{e^{-\beta}}{z_{j\rightarrow i}}
  \sum\limits_{\bm{c}_{\partial j\backslash i}} \biggl[
    \Theta\bigl( \sum\limits_{k\in\partial j\backslash i} c_k\! -\! \theta_j \bigr)
    \prod\limits_{k\in \partial j\backslash i} q_{k\rightarrow j}^{c_k,1}
    + \nonumber \\
    & & 
    \delta_{\sum_{k\in \partial j\backslash i} c_k}^{\theta_j - 1} 
    \prod\limits_{k\in \partial j\backslash i} \bigl( \delta_{c_k}^0
    q_{k\rightarrow j}^{0, 1} + \delta_{c_k}^1 t_{k\rightarrow j}^{1,1} \bigr)
    \biggr] \; ,
  \label{eq:sji11}
\end{eqnarray}
where $\delta_{m}^{n} \! =\! 1$ if $m\! =\! n$ and $=\! 0$ if otherwise. For damage cascading with local recruitments we incorporate the blocking effect of a persuaded neighbor into the second term of Eq.~(\ref{eq:sji11}) to get
\begin{eqnarray}
  & & \hspace*{-0.5cm}
  t_{j\rightarrow i}^{1,1}  =   \frac{e^{-\beta}}{z_{j\rightarrow i}} 
  \sum\limits_{\bm{c}_{\partial j\backslash i}} \biggl[
    \Theta\bigl(\sum\limits_{k\in\partial j\backslash i} c_k \! -\! \theta_j \bigr)
    \prod\limits_{k\in \partial j\backslash i} q_{k\rightarrow j}^{c_k,1} +
    \nonumber \\
    & &
    \delta_{\sum_{k\in \partial j\backslash i} c_k}^{\theta_j -1}
    \Bigl( \prod\limits_{k\in \partial j\backslash i} q_{k\rightarrow j}^{c_k, 1}
    - \prod\limits_{k\in \partial j\backslash i} \bigl( \delta_{c_k}^0
    u_{k\rightarrow j}^{0, 1} + \delta_{c_k}^1 q_{k\rightarrow j}^{1, 1} \bigr)
    \nonumber \\
    & &  \hspace*{2.5cm} +
    \prod\limits_{k\in \partial j\backslash i} \bigl( \delta_{c_k}^0
    u_{k\rightarrow j}^{0, 1} + \delta_{c_k}^1 t_{k\rightarrow j}^{1,1} \bigr)
    \Bigr)
    \biggr] \; .
  \label{eq:sji11modify2}
\end{eqnarray}
To appreciate this modification, notice that node $j$ will not flip if it has a persuadable inactive neighbor~\cite{SInote2019}.

With these preparations we can now express the marginal probability $t_i$ of node $i$ being active but \emph{not} being a break node as
\begin{equation}
  t_i \! =\! \frac{e^{-\beta} \sum\limits_{\bm{c}_{\partial i}}
    \Theta\bigl(\sum\limits_{j\in\partial i} c_j\! -\! \theta_i \bigr)
    \prod\limits_{j\in\partial i} \bigl( \delta_{c_j}^0 q_{j\rightarrow i}^{0,1}
    + \delta_{c_j}^1 t_{j\rightarrow i}^{1, 1} \bigr)}{
    \sum\limits_{\bm{c}_{\partial i}} \Bigl[
      \prod\limits_{j\in \partial i} q_{j\rightarrow i}^{c_j, 0} + e^{-\beta}
      \Theta\bigl(\sum\limits_{j\in\partial i} c_j \! -\! \theta_i \bigr)
      \prod\limits_{j\in\partial i} q_{j\rightarrow i}^{c_j,1}  \Bigr]
  } \; ,
  \label{eq:ti}
\end{equation}
where $\bm{c}_{\partial i} \!\equiv\!\{c_j \!:\! j\!\in\! \partial i\}$ is a composite state of node $i$'s neighbors. The marginal probability $q_i$ of node $i$ being active has the same expression as Eq.~(\ref{eq:ti}) but with $t_{j\rightarrow i}^{1,1}$ replaced by $q_{j\rightarrow i}^{1,1}$~\cite{Xu-etal-2018}. The average fractions of break nodes in the damage cascading process without and with local recruitments are computed by the same expression $\phi\, (\rm{and}\,\, \psi) = \sum_{i=1}^{N} (q_i \! -\! t_i)/N$, with $t_{j\rightarrow i}^{1,1}$ in Eq.~(\ref{eq:ti}) fixed by Eq.~(\ref{eq:sji11}) and Eq.~(\ref{eq:sji11modify2}), respectively. We work on the microcanonical ensemble of fixed energy density $\rho$, so the inverse temperature $\beta$ is determined by the energy constraint $\rho = \sum_{i=1}^{N} q_i /N$.

\emph{Simulation method.} To check the predictions of our mean field theory, we adopt the demon algorithm of microcanonical Monte Carlo (MMC) simulation to sample alliance configurations with equal weight~\cite{Creutz-1983,Rose-Machta-2019}. At each elementary MMC step a new alliance configuration $\bm{c}^\prime$ is proposed by flipping under detailed balance a single node or a tree of same-state nodes of the incumbent $\bm{c}$~\cite{Xu-etal-2018,SInote2019}. If the energy of $\bm{c}^\prime$ does not exceed an objective value $E_o\! \equiv\! \rho N$ then $\bm{c}^\prime$ is accepted as the next configuration of the network, otherwise the network adheres to $\bm{c}$. One unit time of this MMC dynamics corresponds to $N$ consecutive trials of configuration transitions. At each energy density $\rho$ we typically collect $10^5$ configurations at unit time interval~\cite{SInote2019}.

\begin{figure*}[t]
  \centering
    \subfigure[]{
    \label{fig:RRD6:a}
    \includegraphics[angle=270,width=0.3\linewidth]{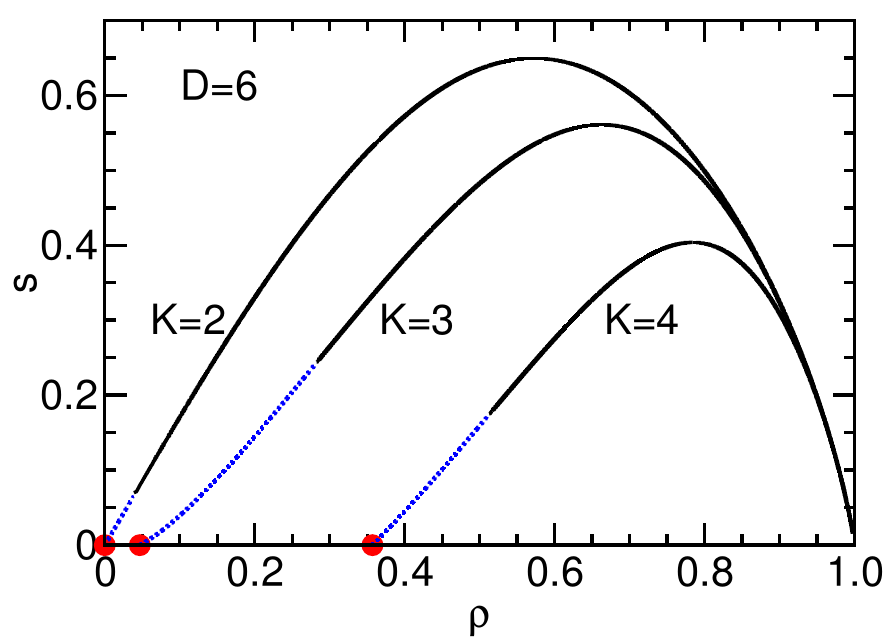}}
  \subfigure[]{
    \label{fig:RRD6:b}
    \includegraphics[angle=270,width=0.3\linewidth]{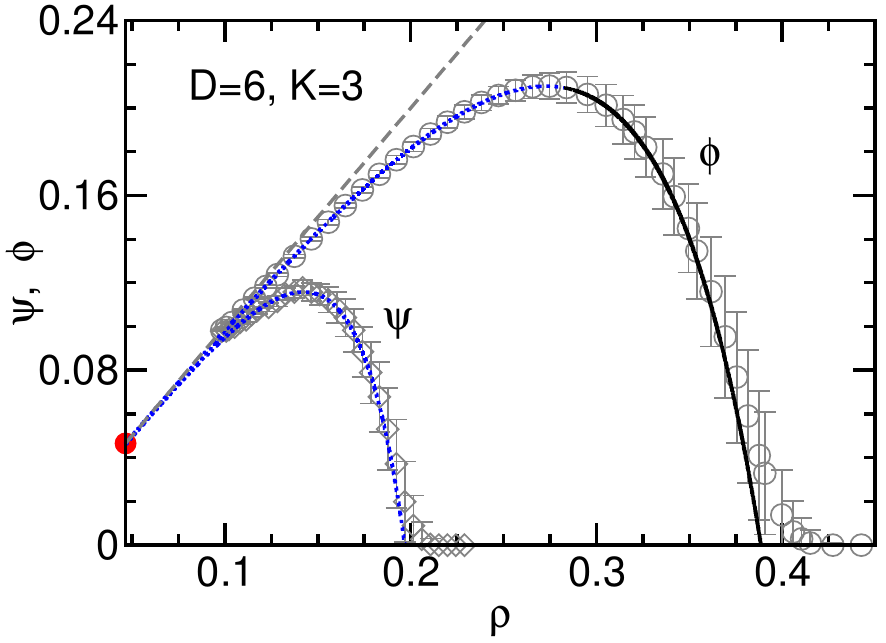}}
  \subfigure[]{
    \label{fig:RRD6:c}
    \includegraphics[angle=270,width=0.3\linewidth]{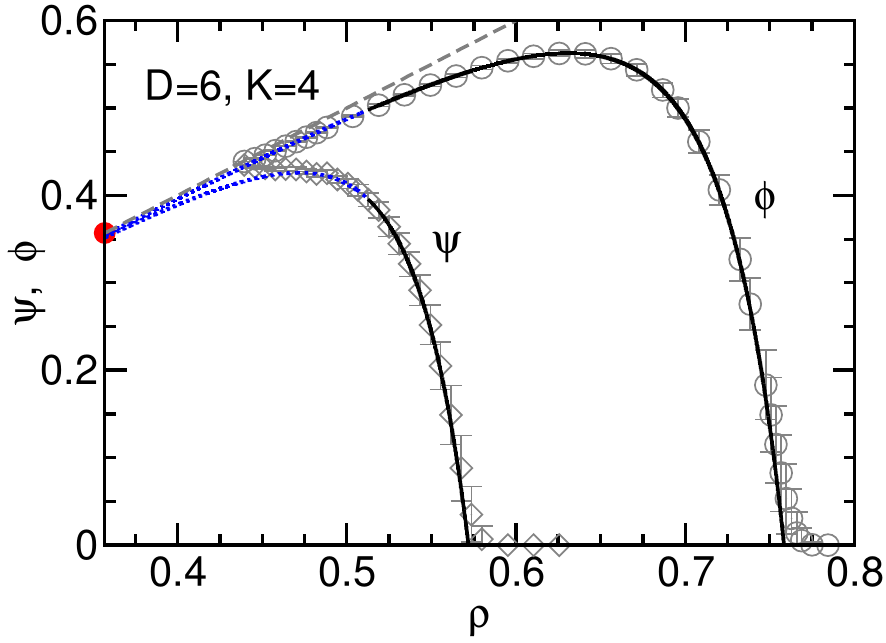}}

  \subfigure[]{
    \label{fig:RRD6:d}
   \includegraphics[angle=270,width=0.3\linewidth]{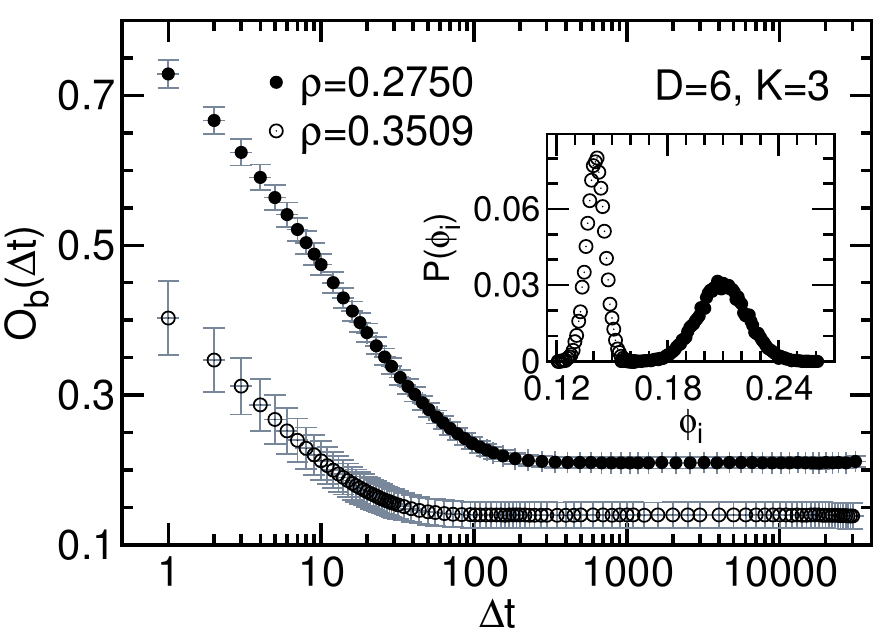}}
  \subfigure[]{
    \label{fig:RRD6:e}
    \includegraphics[angle=270,width=0.3\linewidth]{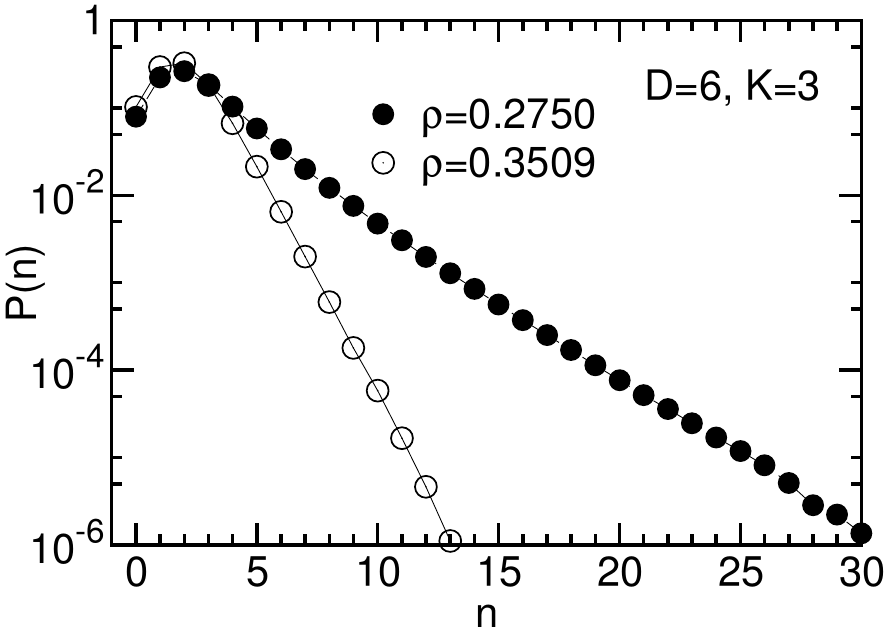}}
  \caption{
    \label{fig:RRD6}
    The RR network ensemble of degree $D\! = \! 6$ and uniform threshold $K$. (a): Entropy density $s$ versus energy density $\rho$ ($K\! = \! 2, 3, 4$). Dotted and solid lines in (a)-(c) indicate respectively the convex and concave branch of $s(\rho)$, and filled circles mark the minimum energy density $\rho_0$. (b) and (c): Density of break nodes without ($\phi$) or with ($\psi$) local recruitments at $K\! = \! 3$ (b) and $K\! = \! 4$ (c). Dashed lines denote the upper bound $\phi\! = \! \rho$. (d): Overlap of break nodes, $O_{\rm{b}}(\Delta t) \equiv \frac{2 | B(\bm{c})\, \cap\, B(\bm{c}^\prime)|}{| B(\bm{c}| + |B(\bm{c}^\prime)|}$, between two alliance configurations $\bm{c}$ and $\bm{c}^\prime$ sampled at time interval $\Delta t$, where $B(\bm{c})$ is the set of break nodes in configuration $\bm{c}$ ($K\! = \! 3$). The inset shows the distribution of $\phi_i$ for all the nodes, where $\phi_i$ is the frequency of node $i$ being a break node in a sampled configuration. (e): Exponentially decaying distribution $P(n)$ of the cost $n$ to suppress a global collapse, where $n$ is the number of recruited persuadable nodes during a damage cascading process with local recruitments ($K\! = \! 3$).  The relative size $\rho \! = \! 0.2750$ or $0.3509$ in (d) and (e). Symbols in (b)-(e) are simulation results obtained on a single network of size $N\! = \! 32768$, and error bars denote standard deviations.
  }
\end{figure*}

\section{Results}

\emph{Uniform threshold $\theta_i \! \equiv \! K$.} The damage cascading process would never exponentially proliferate if $\theta_i \! \leq \! 2$ for all the nodes and then catastrophic collapses would never occur~\cite{SInote2019}. We therefore consider the nontrivial situations that at least some nodes $i$ have $\theta_i \! \geq \! 3$. First, we study the situation of uniform threshold ($\theta_i \! \equiv \! K$). An alliance is then equivalent to a $K$-core~\cite{Chalupa-Leath-Reich-1979,Pittel-etal-1996,Dorogovtsev-etal-2006,Schwarz-Liu-Chayes-2006}. We examine regular random (RR) networks in which every node has exactly $D$ neighbors~\cite{Albert-Barabasi-2002}, and for which the mean-field equations are much simplified because $q_{j\rightarrow i}^{c_j, c_i}$ and $t_{j\rightarrow i}^{1,1}$ are the same for all the links~\cite{Xu-etal-2018,SInote2019}.

The theoretical results for $D\! = \! 6$ and $K\! \in \!\{2, 3, 4\}$ are shown in Fig.~\ref{fig:RRD6}. The entropy density $s(\rho)$ is convex for $\rho\! < \! \rho_{\rm{x}}$ and concave for $\rho \! > \! \rho_{\rm{x}}$, where $\rho_{\rm{x}}$ is the inflection point [Fig.~\ref{fig:RRD6:a}]. The convexity of $s(\rho)$ indicates that alliances of relative sizes $\rho \! < \! \rho_{\rm{x}}$ are difficult to construct~\cite{Xu-etal-2018}. For $K \! \geq \! 3$ and without local protective recruitments, we indeed find that the fraction $\phi$ of break nodes becomes positive as $\rho$ decreases below a critical value $\rho_{\rm{wt}}$ (the weak tipping point) which is considerably larger than $\rho_{\rm{x}}$. With local recruitments, however, the break-node fraction $\psi$ is zero as long as $\rho \! > \! \rho_{\rm{st}}$ with $\rho_{\rm{st}}$ (the strong tipping point) being much smaller than $\rho_{\rm{wt}}$. At $D\! = \! 6$ we have $\rho_{\rm{wt}}\! = \! 0.3885$, $\rho_{\rm{st}}\! =\! 0.1968$ for $K\!= \! 3$ [Fig.~\ref{fig:RRD6:b}] and $\rho_{\rm{wt}}\! = \! 0.7585$, $\rho_{\rm{st}}\! =\! 0.5719$ for $K\!= \! 4$ [Fig.~\ref{fig:RRD6:c}]. The strong tipping point is in the entropy-convex region ($\rho_{\rm{st}} \! < \! \rho_{\rm{x}}$) for $K\! =\! 3$ but it is in the entropy-concave region ($\rho_{\rm{st}} \! > \! \rho_{\rm{x}}$) for $K\! \geq \! 4$.

The predicted relationships of $\phi$, $\psi$ with $\rho$ are confirmed by our MMC results when the ratio $\phi/\rho$ is not too close to unity (Fig.~\ref{fig:RRD6}). (When $\phi$ approaches $\rho$ it becomes exceedingly hard to equilibrate the MMC dynamics.) When $\rho$ is only slightly below the respective phase transition value $\rho_{\rm wt}$ and $\rho_{\rm st}$, we find that
\begin{equation}
  \label{eq:scaling}
  \phi \propto (\rho_{\rm wt} - \rho)^\zeta
  \; , \quad \quad 
  \psi \propto (\rho_{\rm st} - \rho)^\zeta \; ,
\end{equation}
with scaling exponent $\zeta = 1$. These critical scaling relationships are also observed in other network ensembles.

Although at $\rho\! < \! \rho_{\rm{wt}}$ some of the active nodes can be distinguished as break nodes, we find that this heterogeneity is not static but quite dynamic: the set $B(\bm{c})$ of break nodes changes quickly as the equilibrium alliance configuration $\bm{c}$ evolves with time and its relative size fluctuates around $\phi$ to some extent, and the chance of being a break node is similar for all the nodes of the RR network (Fig.~\ref{fig:RRD6:d}, dynamical heterogeneity). If a catastrophic avalanche is suppressible by local recruitments, we find that the number of actually recruited nodes during the whole cascading process is only of order unity [Fig.~\ref{fig:RRD6:e}], confirming that the local recruitment intervention is a minimum-cost strategy.

\begin{figure}[t]
  \centering
    \subfigure[]{
    \label{fig:RRD6K3:a}
     \includegraphics[angle=270,width=0.469\linewidth]{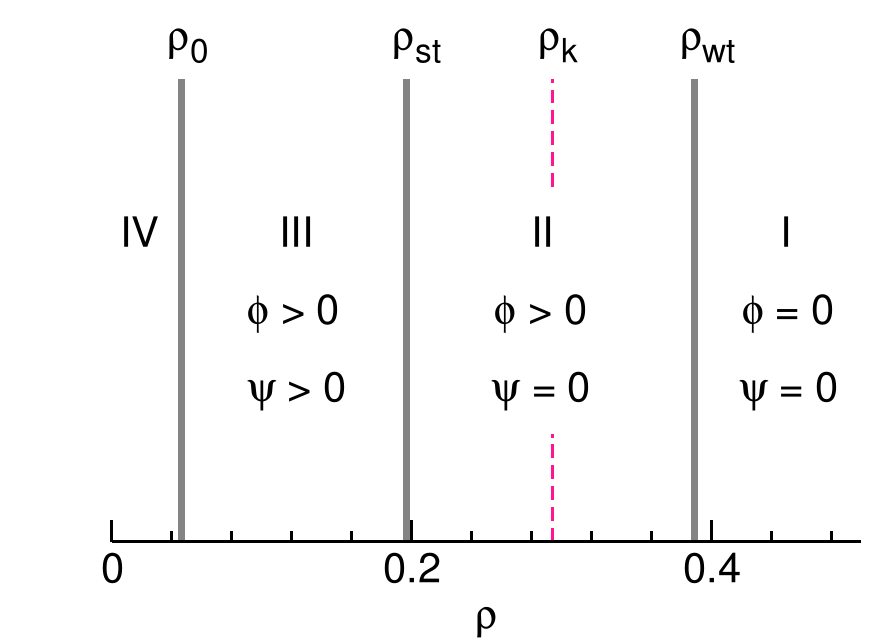}}
  \subfigure[]{
    \label{fig:RRD6K3:b}
    \includegraphics[angle=270,width=0.469\linewidth]{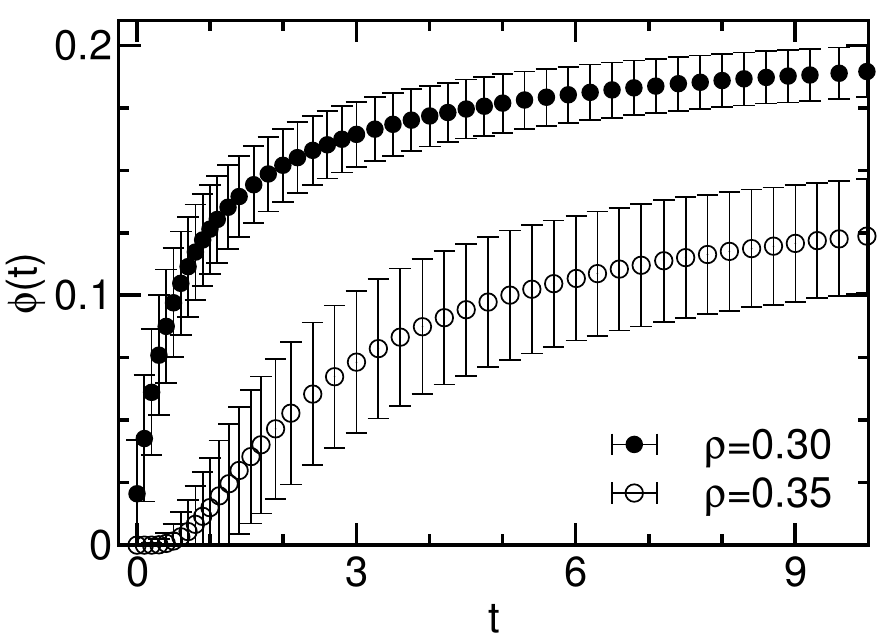}}
  \caption{
    \label{fig:RRD6K3}
    Equilibrium alliance configurations are more vulnerable than kinetic ones. (a): Phase diagram for the RR network ensemble of degree $D\! = \! 6$ and threshold $K \! = \! 3$; $\rho_{\rm wt}$ and $\rho_{\rm st}$ are the weak and strong and tipping points, $\rho_{\rm k}$ and $\rho_{0}$ are the minimum energy density of kinetic and equilibrium alliance configurations. (b): Evolution of break-node fraction $\phi$ with time $t$ at fixed energy density $\rho\! =\! 0.30$ or $0.35$ starting from initial kinetic alliance configurations. Each data point shows the mean and standard deviation of $\phi$ over $48000$ independent trajectories obtained on a single RR network of size $N \! = \! 32768$, $D\! = \! 6$ and $K\! = \! 3$.
    }
\end{figure}

We can summarize these results by a phase diagram of equilibrium alliances [Fig.~\ref{fig:RRD6K3:a}]. When energy density $\rho \! > \! \rho_{\rm wt}$ the system is robust to local perturbations ($\phi \! = \! \psi \! = 0$, phase I); a continuous phase transition occurs at $\rho_{\rm wt}$ and the system then becomes vulnerable to local perturbations but it remains resilient due to the local protection mechanism ($\phi \! > \! 0$ and $\psi \! = \! 0$, phase II); a second continuous phase transition occurs when $\rho$ decreases to $\rho_{\rm st}$, at which the system is no longer resilient ($\phi \! > \! 0$ and $\psi \! > \! 0$, phase III); and finally no equilibrium alliance configurations is possible when $\rho$ decreases below the minimum value $\rho_0$ (phase IV, the forbidden region).

By the irreversible $K$-core pruning process starting from completely random activity patterns, we can reach kinetic alliance configurations of relative size $\rho$ much smaller than $\rho_{\rm wt}$ and down to the kinetic threshold value $\rho_{\rm{k}}$ ($= \! 0.2942$ for $K\! = \! 3$ and $= \! 0.6574$ for $K\! = \! 4$, at degree $D \! = \! 6$)~\cite{Goltsev-etal-2006,SInote2019}. In the thermodynamic limit of $N\! \rightarrow \! \infty$ all such kinetic alliance configurations are robust against single-node perturbations ($\phi \! = 0$), which is reasonable because otherwise they can not survive the pruning process~\cite{SInote2019}. The prediction $\rho_{\rm wt} \! > \! \rho_{\rm k}$ indicates that equilibrium alliance configurations are more vulnerable to local perturbations than are kinetic ones [Fig.~\ref{fig:RRD6K3:a}]. Indeed if kinetic alliance configurations of relative size $\rho\! \in\! (\rho_{\rm{k}}, \rho_{\rm{wt}})$ are allowed to relax towards equilibrium the fraction $\phi$ of break nodes will increase from zero and gradually approach the equilibrium value [Fig.~\ref{fig:RRD6K3:b}]. This counter-intuitive buildup of fragility is driven by entropy maximization. It highlights the fact that kinetic alliances are only rare atypical configurations whose properties are far from being representative of those typical (equilibrium) alliance configurations at the same energy density.

Besides RR networks, we also apply our theory and MMC algorithm to several other types of networks with narrowly distributed degrees, including cubic lattices, small-world networks~\cite{Watts-Strogatz-1998}, Erd\"os-R\'enyi (ER) networks~\cite{Albert-Barabasi-2002}, and peer-to-peer computer server networks~\cite{Ripeanu-etal-2002}. Results obtained on these networks are quite similar to Fig.~\ref{fig:RRD6:b} and \ref{fig:RRD6:c}, see Ref.~\cite{SInote2019} for details.

\emph{Degree-dependent thresholds.}  If the network has a broad degree profile, e.g. scale-free (SF) ~\cite{Leskovec-Kleinberg-Faloutsos-2007,Goh-Kahng-Kim-2001}, while the thresholds $\theta_i$ still remain uniform, the equilibrium alliance configurations will always be robust to random perturbations and $\phi \! = \! 0$~\cite{SInote2019}. This super-robustness is attributed to the high-degree hub nodes which have a strong stabilizing effect~\cite{Dorogovtsev-etal-2006}. For such systems it may be more appropriate to assume that the threshold $\theta_i$ of node $i$ will be higher if it has a larger degree $d_i$. To be concrete, here we assume the linear relationship $\theta_i = r d_i$. Here the parameter $r \in [0, 1]$ denotes the required minimum fraction of active neighbors by an active node~\cite{Watts-2002}.

\begin{figure}
  \centering
  \subfigure[]{
    \label{fig:ERf0p55BPa}
    \includegraphics[angle=270,width=0.469\linewidth]{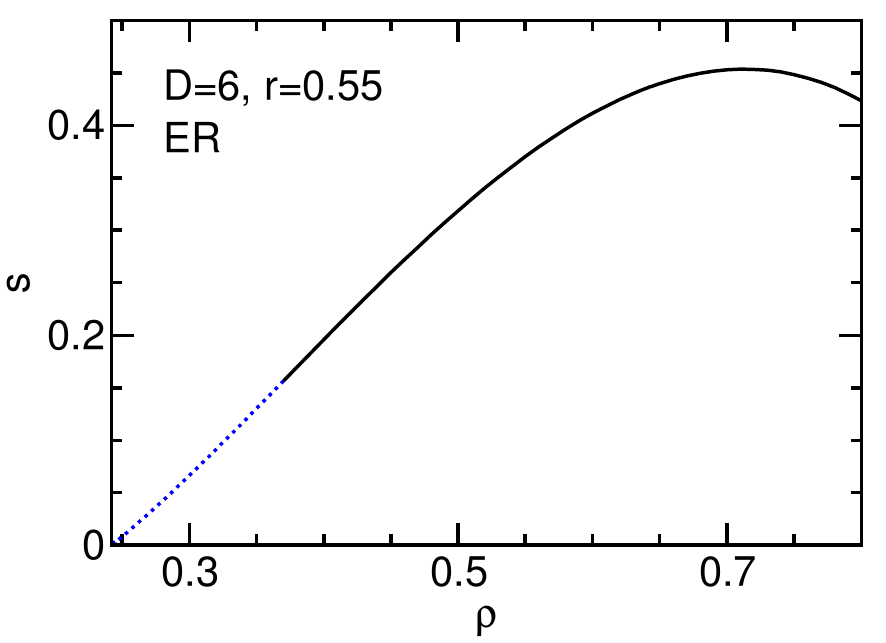}}
  \subfigure[]{
    \label{fig:ERf0p55BPb}
    \includegraphics[angle=270,width=0.469\linewidth]{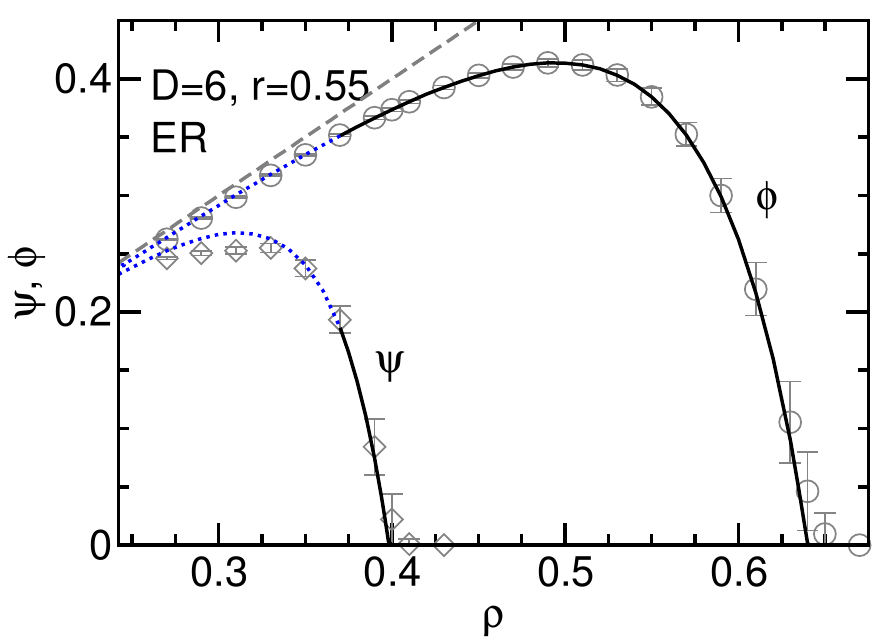}}

  \subfigure[]{
    \label{fig:SFf0p55BPa}
    \includegraphics[angle=270,width=0.469\linewidth]{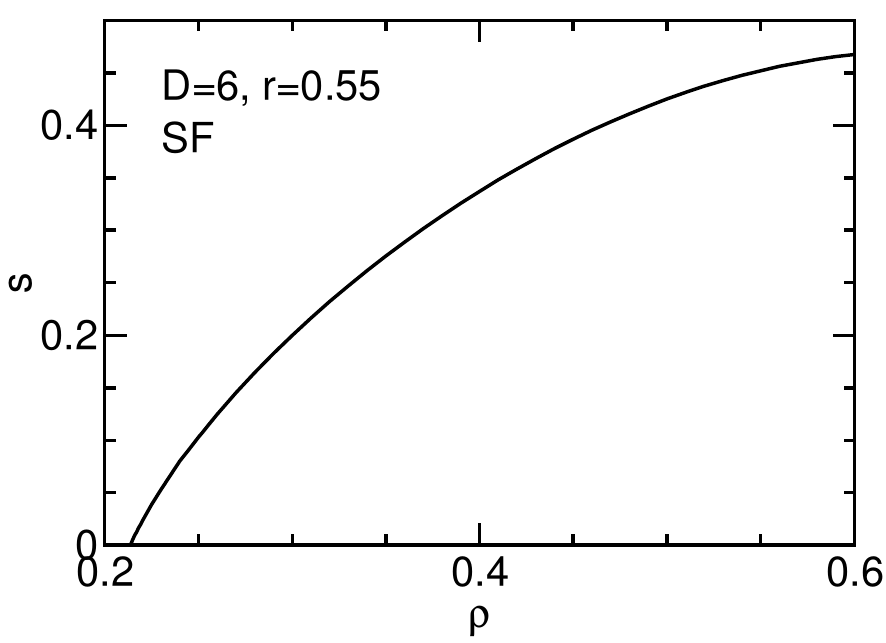}}
  \subfigure[]{
    \label{fig:SFf0p55BPb}
    \includegraphics[angle=270,width=0.469\linewidth]{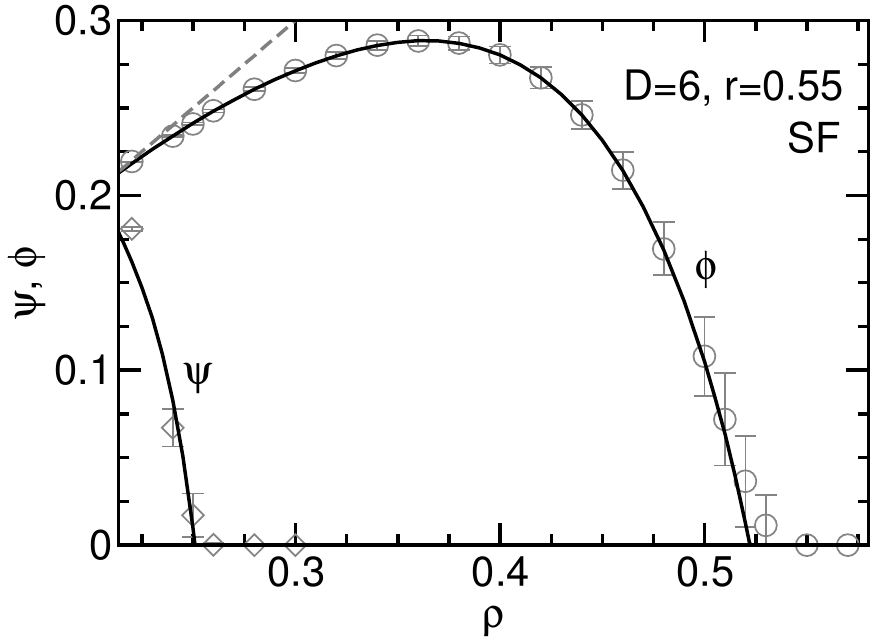}}
  \caption{
    \label{fig:PropModel0p55}
    Results on equilibrium alliances of ER (a, b) and SF (c, d) random networks of mean degree $D \! = \! 6$ and degree-dependent thresholds $\theta_i \! = \! r d_i$ with $r\! = \! 0.55$.  (a, c): Entropy density $s$ versus energy density $\rho$. (b, d): Density of break nodes with ($\phi$) and with ($\psi$) local recruitments. Lines are theoretical predictions; dotted and solid lines indicate respectively the convex and concave branch of $s(\rho)$. Symbols are simulation results obtained on a single network instance of size $N \! = \! 40000$; error bars denote standard deviations. The SF network instance is constructed following the static method~\cite{Goh-Kahng-Kim-2001} and its degree power-law exponent is $\gamma \! = \! 3.0$.
    }
\end{figure}

Some representative results obtained on ER and SF network instances \revision{(with $r = 0.55$)} are reported in Fig.~\ref{fig:PropModel0p55}, demonstrating the characteristic phenomenon of two distinct dynamical phase transitions and the scaling behavior (\ref{eq:scaling}). The good agreement between theory and simulation also confirms the wide applicability of our mean-field theory to different network types.  Figure~\ref{fig:PropModel0p55} reveals that for SF and ER networks with the same mean degree value $D$, the critical energy densities $\rho_{\rm wt}$ and $\rho_{\rm st}$ are shifted to lower values in the SF networks as compared to the corresponding values of the ER networks. This indicates that SF networks have enhanced robustness and resilience than the ER networks and therefore are evolutionarily more favorable.

\section{Discussion}

To briefly conclude, we presented a mean field theory for the network equilibrium social engagement and alliance problem and revealed the existence of two continuous dynamical phase transitions. We pointed out that the kinetic alliance configurations widely discussed in the literature are out-of-equilibrium atypical configurations and they may become more and more vulnerable to local perturbations when they relax.

Our theory can be extended in several different ways. First, interactions in many real-world social networks are asymmetric and therefore it is necessary to consider the alliance problem on networks formed by directed and weighted links. Second, the nodes in social networks are often quite heterogenous in their influences to others. The effect of influence heterogeneity can be investigated as a node-weighted alliance problem. Alliance configurations are a type of relatively rigid structures in a network, and as such they are closely related to jammed states in granular systems with similar properties such as dynamical heterogeneity and rigidity percolation~\cite{Schwarz-Liu-Chayes-2006,Yang-etal-2019}. It is interesting to refine the alliance model into a more realistic network jamming model for systems embedded in three dimensional spaces.

Our work revealed that localized intervention measures have a huge effect on the stability of a complex system, shifting the tipping point $\rho_{\rm{wt}}$ to a much smaller value $\rho_{\rm{st}}$. The proposed local recruitment strategy may not be optimal. For example, by enlarging the length scale of local interventions it may be possible to further reduce the value of $\rho_{\rm{st}}$. The effect of other types of protection strategies also deserves study. Besides such passive damage responses, a complex system formed many mutually dependent ``cells'' may also actively exercise itself to improve and rejuvenate its functions. This is another interesting issue for future theoretical investigations.

%\begin{acknowledgments}
\acknowledgments
This work was supported by the National Natural Science Foundation of China Grants No.11975295 and No.11947302, and the Chinese Academy of Sciences Grant No.QYZDJ-SSW-SYS018. Numerical simulations were carried out at the Tianwen and HPC clusters of ITP-CAS. S.N.W. and L.C. contributed equally to this work. Corresponding authors: H.J.Z. (zhouhj@itp.ac.cn) and L.C. (luancheng@dlut.edu.cn).
%\end{acknowledgments}

%\bibliography{/Users/zhouhj/references}

\clearpage

\begin{center}
  {\bf{Vulnerability and Resilience of Social Engagement: Equilibrium Theory}}
  \vskip 0.1cm
  Shang-Nan Wang, Luan Cheng, Hai-Jun Zhou
  \vskip 0.3cm
  Supplementary Information
\end{center}

\vskip 0.2cm

%\tableofcontents

\section{Mean field theory for the equilibrium alliance problem}
\label{sec:mfea}

The mean field theory for the strong defensive alliance problem, described in Ref.~\cite{Xu-etal-2018}, is also applicable to the slightly more general alliance problem studied in the present work. Here we review this theory as a warm-up to the theory of alliance collapse, developed in the next section.

\subsection{Partition function and factor graph representation}

For the alliance problem defined on a network $G$ of $N$ nodes and $M$ links between these nodes, the partition function is
\begin{equation}
  \label{eq:Zbeta}
  Z(\beta)=\sum\limits_{\bm{c} \neq \bm{0}} \prod\limits_{i=1}^{N}
  \biggl[ e^{-\beta c_i} \Theta\Bigl (\sum_{j\in \partial i} c_j
    - \theta_i c_i  \Bigr)\biggr] \; .
\end{equation}
In this expression, $\bm{c} \equiv (c_1, c_2, \ldots, c_N)$ with $c_i \in \{0, 1\}$ denotes a generic activity configuration of the $N$ nodes, and $\bm{0}\equiv (0, 0, \ldots, 0)$ is the completely inactive configuration; $\beta$ is the inverse temperature parameter; $\Theta(x)$ is the Heaviside step function such that $\Theta(x)\! =\! 0$ for $x\! <\! 0$ and $\Theta(x)\! =\! 1$ for $x\! \geq \! 0$; $\partial i$ denotes the node set formed by all the nearest neighbors of node $i$ in the network; and $\theta_i$ is the threshold parameter for node $i$. Denote the set of active nodes in configuration $\bm{c}$ as $A(\bm{c})$, that is, $A(\bm{c}) \equiv \{ i:  c_i = 1\}$. If the condition $\sum_{j\in \partial i} c_j \geq \theta_i$ holds for every node $i\in A(\bm{c})$ and therefore $A(\bm{c})$ is a valid alliance,  then $\bm{c}$ contributes a term $e^{- \beta E(\bm{c})}$ to the partition function, where $E(\bm{c}) \equiv \sum_{i=1}^N c_i$ is the energy of $\bm{c}$ (the cardinality of set $A(\bm{c})$). On the other hand if $\sum_{j\in \partial i} c_j$ is less than $\theta_i$ for at least one node $i\in A(\bm{c})$, then $A(\bm{c})$ is not a valid alliance and the contribution of $\bm{c}$ to $Z(\beta)$ is exactly zero. Therefore, although $Z(\beta)$ is a sum over all the $2^N$ binary configurations except for $\bm{0}$, only the valid alliance configurations have positive contribution to $Z(\beta)$.

As the form of Eq.~(\ref{eq:Zbeta}) suggests, there are $N$ many-body interactions in the system, each of which is brought by a node (say $i$) and it involves $i$ and all its nearest neighbors $j$. The Boltzmann weight of such an interaction is $\Psi_i(c_i, \bm{c}_{\partial i}) \equiv e^{-\beta c_i}  \Theta\bigl (\sum_{j\in \partial i} c_j - \theta_i c_i  \bigr)$, where $\bm{c}_{\partial i} \equiv \{c_j : j\in \partial i\}$ denotes an activity pattern of the nodes in the set $\partial i$, and the partition function is then $Z(\beta) = \sum_{\bm{c}\neq \bm{0}} \prod_{i=1}^{N} \Psi_i(c_i, \bm{c}_{\partial i})$.

\begin{figure}[t]
  \centering
  \includegraphics[width=0.618\textwidth]{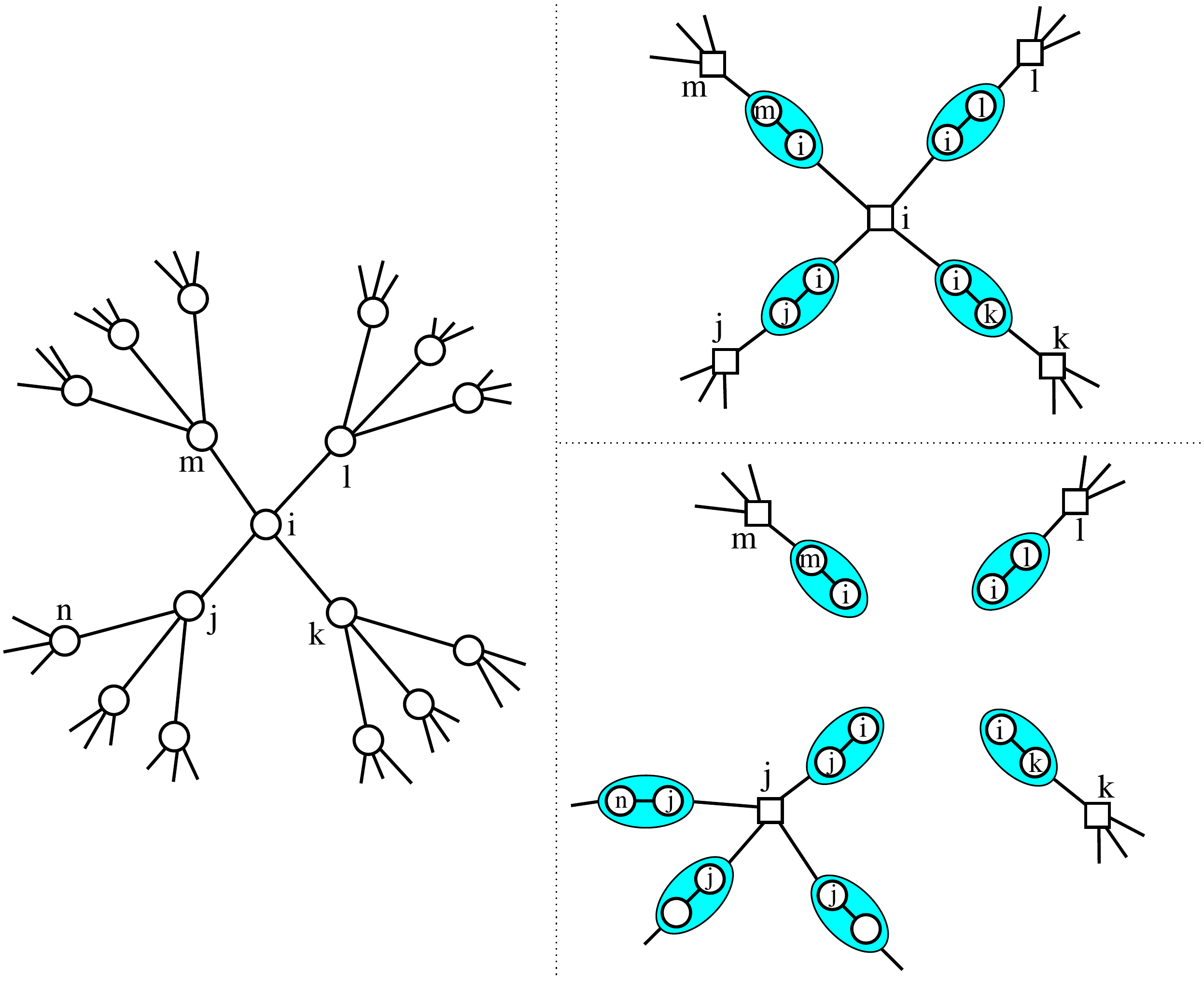}
  \caption{
    \label{fig:fg}
    (left) The neighborhood of a node $i$ in the original network $G$. In this example the degree of every node is $D\! = \! 4$, and the set of nearest neighbors is $\partial i = \{j, k, l, m\}$ for node $i$. Node $n$ is one of the nearest neighbors of node $j$.  (upper right) The neighborhood of the factor-node $i$ in the bipartite factor graph, corresponding to the neighborhood of node $i$ in the original network $G$. A link $(i, j)$ of network $G$ is represented by a variable-node in the factor graph (shown as an ellipse), and is assigned a composite state $(c_i, c_j)$. A many-body interaction involving node $i$ and all its nearest neighbors in network $G$ is represented by a factor-node $i$ in the factor graph (shown as a square). Each variable-node $(i, j)$ in the factor graph is connected to exactly two factor-nodes $i$ and $j$, because it participates only in two many-body interactions.  (lower right) In the cavity factor graph with factor-node $i$ being removed, variable-node $(i, j)$ is affected only by factor-node $j$ and so its composite state $(c_i, c_j)$ depends only on the composite states of the three other variable-nodes (e.g., ellipse $(j, n)$) attached to factor-node $j$.
  }
\end{figure}

Because of the many-body nature of the weight factors $\Psi_i$, it is convenient to represent the system as a bipartite graph of factor-nodes (representing the $N$ many-body interactions, e.g., square nodes in the right panels of Fig.~\ref{fig:fg}) and variable-nodes (representing the $M$ links, e.g., elliptic nodes in the right panels of Fig.~\ref{fig:fg}). Each variable-node of this bipartite factor graph corresponds to a link $(i, j)$ of the original network, and it has a composite state $(c_i, c_j)$ with four possible values.  Each factor-node $i$ of this bipartite graph is associated with an energy $E_i = c_i$, and further more it imposes the following two constraints: (1) the composite states $(c_i, c_j)$ of all the variable-nodes $(i, j)$ that are connected to the factor-node $i$ should have the same value of $c_i$, and (2) if $c_i\! = \! 1$ then the sum of $c_j$ in the composite states $(c_i, c_j)$ of the connected variable-nodes $(i, j)$ must be at least $\theta_i$.

\subsection{The belief propagation equation}

If short loops are very rare in the original network $G$ so that it is locally tree-like, the corresponding factor graph of $G$ will also be locally tree-like. For such a factor graph, let us consider the situation depicted in the lower-right panel of Fig.~\ref{fig:fg}, where the factor-node $i$ is removed from the factor graph and the variable-node $(i, j)$  participates only in the many-body interaction $j$. Let us denote by $q_{j\rightarrow i}^{c_j, c_i}$ the probability distribution of the composite state $(c_j, c_i)$ of variable-node $(j, i)$ in this cavity graph. As a probability distribution, $q_{j\rightarrow i}^{c_j, c_i}$ satisfies the normalization condition $q_{j\rightarrow i}^{0, 0} + q_{j\rightarrow i}^{1, 0} + q_{j\rightarrow i}^{0, 1} + q_{j\rightarrow i}^{1, 1} \equiv 1$.

To derive a self-consistent equation for the cavity probability distribution $q_{j\rightarrow i}^{c_j, c_i}$, we notice that because loops in a locally tree-like factor graph are usually quite long in the limit of large system size $N$,  all the variable-nodes that are connected to factor-node $j$ in the lower-right panel of Fig.~\ref{fig:fg} could be regarded as mutually independent in the absence of factor-node $j$. Under this assumption, then when the Boltzmann weight $\Psi_j$ of factor-node $j$ is considered (but that of factor-node $i$ is not yet considered), the probability distribution of the composite state $(c_j, c_i)$ will be
\begin{subequations}
  \label{eq:BP}
  \begin{align}
    q_{j\rightarrow i}^{0,0}  & \equiv 
    \frac{1}{z_{j\rightarrow i}} \prod_{k\in \partial j\backslash i} 
    \bigl( q_{k\rightarrow j}^{0,0}+q_{k\rightarrow  j}^{1,0} \bigr) \; ,  \\
  q_{j\rightarrow i}^{0,1} & =
  \frac{1}{z_{j\rightarrow i}} \prod_{k\in \partial j\backslash i} 
  \bigl( q_{k\rightarrow j}^{0,0}+q_{k\rightarrow  j}^{1,0} \bigr) \; ,  \\
  q_{j\rightarrow i}^{1,0} & =  \frac{1}{z_{j\rightarrow i}} e^{-\beta}
  \sum_{\bm{c}_{\partial j\backslash i}}
  \Theta\bigl(\sum\limits_{k\in\partial j\backslash i} c_k 
  -\theta_j \bigr) \prod\limits_{k\in \partial j\backslash i}
  q_{k\rightarrow j}^{c_k,1} \; ,  \\
  q_{j\rightarrow i}^{1,1} & =  \frac{1}{z_{j\rightarrow i}} e^{-\beta}
  \sum\limits_{\bm{c}_{\partial j\backslash i}} 
  \Theta\bigl(\sum\limits_{k\in\partial j\backslash i} c_k + 1 
  -\theta_j \bigr) \prod\limits_{k\in \partial j\backslash i} 
  q_{k\rightarrow j}^{c_k,1} \; ,
  \label{eq:BP:d}
  \end{align}
\end{subequations}
where the set $\partial j\backslash i$ contains all the nearest neighbors of node $j$ except for $i$, and $\bm{c}_{\partial j\backslash i} \!\equiv\! \{c_k \!:\!  k \!\in\! \partial j\backslash i\}$;  $z_{j \rightarrow i}$ is the probability normalization constant:
\begin{eqnarray}
  z_{j\rightarrow i} &  \equiv &  2 \prod_{k\in \partial j\backslash i} 
  \bigl( q_{k\rightarrow j}^{0,0}+q_{k\rightarrow  j}^{1,0} \bigr)
  +  e^{-\beta} \sum_{\bm{c}_{\partial j\backslash i}}
  \Theta\bigl(\sum\limits_{k\in\partial j\backslash i} c_k 
  -\theta_j \bigr) \prod\limits_{k\in \partial j\backslash i}
  q_{k\rightarrow j}^{c_k,1} \nonumber \\
  & & \quad \quad + e^{-\beta}
  \sum\limits_{\bm{c}_{\partial j\backslash i}} 
  \Theta\bigl(\sum\limits_{k\in\partial j\backslash i} c_k + 1 
  -\theta_j \bigr) \prod\limits_{k\in \partial j\backslash i} 
  q_{k\rightarrow j}^{c_k,1} \; .
  \label{eq:zjtoi}
\end{eqnarray}

Equation (\ref{eq:BP}) is the belief-propagation (BP) equation for the alliance problem, applicable to a single network $G$. In the simplest case of regular random (RR) networks $G$ of degree $D$ and uniform threshold parameters $\theta_i \equiv K$ for all the nodes of network $G$, we may further assume that the cavity probability distributions $q_{j\rightarrow i}^{c_j, c_i}$ are the same for all the links $(j, i)$ of the network, i.e., $q_{j\rightarrow i}^{0, 0} = q^{0, 0}$, $q_{j\rightarrow i}^{0,1} = q^{0, 1}$, $q_{j\rightarrow i}^{1, 0} = q^{1, 0}$, and $q_{j\rightarrow i}^{1,1} = q^{1,1}$. Then the BP equation (\ref{eq:BP}) is simplified to
\begin{subequations}
  \label{eq:BPsimple}
  \begin{align}
    q^{0,0} & = 
    \frac{1}{z} \bigl(q^{0,0}+q^{1,0} \bigr)^{D-1} \; ,
    \label{eq:BPsimple00}
    \\
    q^{0,1} & = 
    \frac{1}{z} \bigl(q^{0,0}+q^{1,0} \bigr)^{D-1} \; ,
    \label{eq:BPsimple01}
    \\
    q^{1,0} & =  \frac{1}{z} e^{-\beta}
    \sum\limits_{d\geq K}^{D-1} 
    C_{D-1}^d (q^{1,1})^d (q^{0,1})^{D-1-d} \; ,
    \\
    q^{1,1} & =  \frac{1}{z} e^{-\beta} 
    \sum\limits_{d\geq K-1}^{D-1}C_{D-1}^d
    (q^{1,1})^d (q^{0,1})^{D-1-d} \; ,
  \end{align}
\end{subequations}
where $C_{n}^{m} \equiv \frac{n!}{m! (n-m)!}$ is the binomial coefficient, and the normalization constant $z$ is
\begin{equation}
  \label{eq:zval}
  z \equiv 2  \bigl(q^{0,0}+q^{1,0} \bigr)^{D-1} +
  2  e^{-\beta} \sum\limits_{d\geq K}^{D-1} 
  C_{D-1}^d (q^{1,1})^d (q^{0,1})^{D-1-d}
  +  e^{-\beta}C_{D-1}^{K-1} (q^{1,1})^{K-1} (q^{0,1})^{D-K} \; .
\end{equation}
Notice that $q^{0,0} \equiv q^{0,1}$ according to Eqs.~(\ref{eq:BPsimple00}) and (\ref{eq:BPsimple01}).

A trivial fixed point of Eq.~(\ref{eq:BPsimple}) is $q^{0,0}=q^{0,1} = \frac{1}{2}$, $q^{1,0}=q^{1,1}=0$, which corresponds to the completely inactive configuration $\bm{0}$ (not an alliance). We are interested in the non-trivial fixed points of Eq.~(\ref{eq:BPsimple}) which correspond to alliance configurations of positive cardinality. All these non-trivial fixed-point solutions of Eq.~(\ref{eq:BPsimple}) can be obtained by a simple numerical code, for any fixed value of $\beta$.

\subsection{Average density of active nodes}

Consider the cavity factor graph in the lower-right panel of Fig.~\ref{fig:fg}. As we pointed out earlier, because of the locally tree-like property of this factor graph, the neighboring variable-nodes of the factor-node $i$ could be regarded as mutually independent when the Boltzmann weight $\Psi_i$ is discarded. When factor-node $i$ is added back to the factor graph as in the upper-right panel of Fig.~\ref{fig:fg}, these neighboring variable-nodes will become strongly correlated, including that the $c_i$ values of these variable-nodes must take the same value. At a given value of inverse temperature $\beta$ the probability $q_i$ of $c_i\! =\! 1$ is then expressed as
\begin{equation}
  \label{eq:qi}
  q_i = \frac{e^{-\beta} \sum\limits_{\bm{c}_{\partial i}}
    \Theta\bigl(\sum\limits_{j\in\partial i} c_j -\theta_i \bigr)
    \prod\limits_{j\in\partial i} q_{j\rightarrow i}^{c_j,1} }
      {\prod\limits_{j\in \partial i}
        (q_{j\rightarrow  i}^{0,0} + q_{j\rightarrow i}^{1,0})
        +
        e^{-\beta} \sum\limits_{\bm{c}_{\partial i}} 
        \Theta\bigl(\sum\limits_{j\in\partial i} c_j - \theta_i \bigr)
        \prod\limits_{j\in\partial i}
        q_{j\rightarrow i}^{c_j,1} 
      } \; ,
\end{equation}
where $q_{j\rightarrow i}^{c_j, c_i}$ is a fixed-point solution of Eq.~(\ref{eq:BP}). The average density $\rho$ of active nodes (equivalently, the average energy density) is simply
\begin{equation}
  \label{eq:rho}
  \rho = \frac{1}{N} \sum\limits_{i=1}^{N} q_i \; .
\end{equation}
In the special case of RR networks of degree $D$ and uniform threshold $K$, all the $q_i$ values are equal and then the expression of $\rho$ is simplified to
\begin{equation}
  \label{eq:rhosimple}
  \rho = \frac{e^{-\beta} \sum\limits_{d\geq K}^{D}
    C_{D}^d (q^{1,1})^d  (q^{0,0})^{D-d}}{(q^{0,0}+q^{1,0})^{D}
    + e^{-\beta}\sum\limits_{d\geq K}^{D} C_{D}^d
    (q^{1,1})^d (q^{0,1})^{D-d}} \; .
\end{equation}

We are interested in the microcanonical ensemble, for which the density $\rho$ of active nodes is fixed instead of the inverse temperature $\beta$. To determine the value of $\beta$ for a fixed value of $\rho$, we can treat Eq.~(\ref{eq:rhosimple}) or Eq.~(\ref{eq:rho}) as an equation for $\beta$ and iterate the value of $\beta$ together with the BP equation (\ref{eq:BP}) or (\ref{eq:BPsimple}) to reach a fixed point.

\subsection{Free energy density and entropy density}

For sparse random networks, the free energy $F \equiv -(1/\beta)\ln Z(\beta)$ of the system can be computed through
\begin{equation}
  \label{eq:free}
  F=\sum\limits_{i=1}^{N} f_{i+\partial i} - \sum\limits_{(i, j)\in G} f_{i j} \; ,
\end{equation}
where $f_{i+\partial i}$ is the contribution of factor-node $i$ and all its attached links (variable-nodes), and $f_{i j}$ is the contribution of a single link $(i, j)$. Because each link $(i, j)$ contributes to both $f_{i+\partial i}$ and $f_{j+\partial j}$ its effect is subtracted once in Eq.~(\ref{eq:free}). The explicit expressions for $f_{i+\partial i}$ and $f_{i j}$ are:
\begin{eqnarray}
  f_{i+\partial i} & =& -\frac{1}{\beta}\ln\Bigl[
    \prod\limits_{j\in \partial i}
    (q_{j\rightarrow  i}^{0,0} + q_{j\rightarrow i}^{1,0}) +
    e^{-\beta} 
    \sum\limits_{\bm{c}_{\partial i}}
    \Theta\bigl(\sum\limits_{j\in\partial i} c_j - \theta_i \bigr)
    \prod\limits_{j\in\partial i} q_{j\rightarrow i}^{c_j,1}
    \Bigr] \; , 
  \label{eq:fiandpi}
  \\
  f_{i j} & = & -\frac{1}{\beta}
  \ln\Bigl[q_{i\rightarrow j}^{0,0} q_{j\rightarrow i}^{0,0} +
    q_{i\rightarrow j}^{1,0} q_{j\rightarrow i}^{0,1} + q_{i\rightarrow j}^{0,1}
    q_{j\rightarrow i}^{1,0} + q_{i\rightarrow j}^{1,1} q_{j\rightarrow i}^{1,1}
    \Bigr] \; .
  \label{eq:fij}
\end{eqnarray}
For the special case of RR networks of degree $D$ and uniform threshold $K$, the free energy density $f \equiv F/N$ of the system is
\begin{equation}
  f =  -\frac{1}{\beta} \ln \Bigl[
    (q^{0,0}+q^{1,0})^{D} + e^{-\beta} \sum\limits_{d\geq K}^D C_{D}^d
    (q^{1,1})^d (q^{0,1})^{D-d} \Bigr]
  +\frac{D}{2 \beta} \ln\Bigl[(q^{0,0})^2 + 2 q^{0,1} q^{1,0} + (q^{1,1})^2 \Bigr]
  \; .  
\end{equation}
The entropy density $s$ of the system is simply
\begin{equation}
  s = (\rho - f) \beta \; .
\end{equation}

\subsection{Some extensions of the alliance model}

Here we briefly mention two important extensions of the basic equilibrium alliance problem. The theoretical framework discussed in this and the next section is applicable to these extended models as well.

The first extension is to consider directed networks. In such a network $G$ each link $(i, j)$ has a direction and it is pointing from  node $i$ to node $j$. Directed links can better represent the directed interactions in a social network (e.g., node $i$ influences node $j$ but is not influenced by $j$). The necessary condition for a node $j$ to be active may then be modified as
\begin{equation}
  \sum\limits_{i \in \partial^{+} j} w_{i j} c_i  \geq \theta_j \; ,
\end{equation}
where $\partial^+ j \equiv \{i: (i, j)\in G\}$ denotes the set of upstream nearest neighbors $i$ of node $j$, and $w_{i j}$ is the weight of link $(i, j)$. In the default case the link weights $w_{i j}$ may all be set to be uniform.

The second extension is to consider undirected (or directed) networks whose nodes $i$ have quite different weights $w_i$. For example, opinion leaders or celebrities in social networks would have a large influence on ordinary people but not the other way around, and these influential nodes should have much larger weights than the other nodes. For a node-weighted undirected network, the necessary condition for a node $j$ to be active may then be modified as
\begin{equation}
  \sum\limits_{i \in \partial j} w_{i} c_i  \geq \theta_j \; .
\end{equation}

\clearpage

\section{Mean field theory of network alliance collapse}
\label{sec:phipsi}

We now derive some mean field iterative equations concerning the stability of an equilibrium alliance configuration $\bm{c}$, sampled uniformly at random from all the alliance configurations with the same density $\rho$ of active nodes.

We classify the inactive nodes of the configuration $\bm{c}$ into two types, persuadable and non-persuadable. An inactive node $j$ of $\bm{c}$ is persuadable if it has $\theta_j$ or more active nearest neighbors (that is, $\sum_{k\in \partial j} c_j \geq \theta_j$), otherwise $j$ is non-persuadable. Notice that a persuadable inactive node can be recruited to the alliance $A(\bm{c})$.

Consider a node (say $i$) which is active ($c_i\! =\! 1$) in configuration $\bm{c}$. If now this node drops out of the alliance $A(\bm{c})$ and then keeps inactive ($c_i\! =\! 0$) indefinitely, it may pull some of its active nearest neighbors down to the inactive state, and these induced changes may again make some of the active second nearest-neighboring nodes to quit $A(\bm{c})$, and so on. For sufficiently large sparse random networks, because of the extreme rareness of short loops (i.e., the locally tree-like property), the whole affected region after the damage cascading process finally stops either forms a tree structure of small size, if the radius of damage is small and independent of network size $N$, or it is a giant loop-rich sub-network involving all (or at least a major fraction, say $>\! 80\%$) of the nodes of the alliance $A(\bm{c})$. In the latter case of extensive damage the radius of the damage region increases with the system size $N$ in a logarithmic manner~\cite{Zhou-2005a} and we regard it as a collapse of the alliance $\bm{A}$. In the following discussions we regard an active node $i$ of $\bm{c}$ as ordinary active if its breakdown only induces a small (tree-formed) subnetwork of damaged nodes. If instead the breakdown of node $i$ induces a giant sub-network of damaged nodes we say that node $i$ is a break node of $A(\bm{c})$.

\subsection{Damage propagation without local recruitments}
\label{sec:mftnr}

We first discuss the simplest scenario of no local recruitments, namely that no any persuadable inactive node of $\bm{c}$ will be recruited to the alliance to help stabilizing it during the damage cascading process initialized from an active node $i$. Consider again the cavity factor graph in the lower-right panel of Fig.~\ref{fig:fg}, in which the factor-node $i$ is removed. Let us denote by $t_{j\rightarrow i}^{1, 1}$ the probability that (1) the composite state $(c_j, c_i)$ of the variable-node $(j, i)$ is $(1, 1)$ in this cavity system and (2) if $c_i$ changes to $c_i\! =\! 0$ it will induce a small sub-network of perturbed nodes.  Because short loops are absent in the random factor graph, the induced sub-network is a tree. Then we obtain that
\begin{equation}
  t_{j\rightarrow i}^{1,1} =   \frac{e^{-\beta}}{z_{j\rightarrow i}} 
  \sum\limits_{\bm{c}_{\partial j\backslash i}} \Bigl[
    \Theta\bigl( \sum\limits_{k\in\partial j\backslash i} c_k  -\theta_j \bigr)
    \prod\limits_{k\in \partial j\backslash i} q_{k\rightarrow j}^{c_k,1}
    + \delta_{\sum_{k\in \partial j\backslash i} c_k}^{\theta_j -1}
    \prod\limits_{k\in \partial j\backslash i} \bigl( \delta_{c_k}^0
    q_{k\rightarrow j}^{0, 1} + \delta_{c_k}^1 t_{k\rightarrow j}^{1,1} \bigr)
    \Bigr] \; ,
  \label{eq:sji11}
\end{equation}
where $z_{j\rightarrow i}$ is computed through Eq.~(\ref{eq:zjtoi}). The first term within the square brackets corresponds to the situation of node $j$ having more than $\theta_j$ active nearest neighbors ($j$ will not jump to $c_j\! =\! 0$ when $c_i$ jumps to $c_i\! =\! 0$); the second term corresponds to the situation of node $j$ having exactly $\theta_j$ active nearest neighbors ($j$ will jump to $c_j\! =\! 0$ when $c_i$ jumps to $c_i\!=\! 0$).

Similar to the above discussion, in the case of node $i$ being connected to all its nearest neighbors (see the upper-right panel of Fig.~\ref{fig:fg}), the probability $t_{i}$ of node $i$ being an ordinary active node (not being a break node) is expressed as
\begin{equation}
  t_i = \frac{e^{-\beta} \sum\limits_{\bm{c}_{\partial i}}
    \Theta\bigl(\sum\limits_{j\in\partial i} c_j -\theta_i \bigr)
    \prod\limits_{j\in\partial i} \bigl( \delta_{c_j}^0 q_{j\rightarrow i}^{0,1}
    + \delta_{c_j}^1 t_{j\rightarrow i}^{1, 1} \bigr)}{
    \prod\limits_{j\in \partial i}
    (q_{j\rightarrow  i}^{0,0} + q_{j\rightarrow i}^{1,0})
    + e^{-\beta} \sum\limits_{\bm{c}_{\partial i}} 
    \Theta\bigl(\sum\limits_{j\in\partial i} c_j - \theta_i \bigr)
    \prod\limits_{j\in\partial i}
    q_{j\rightarrow i}^{c_j,1}  
  } \; .
\end{equation}
Because an active node is either ordinarily active or is a break node,  the mean fraction $\phi$ of break nodes in the network is then computed through
\begin{equation}
  \phi \equiv \frac{1}{N} \sum\limits_{i=1}^{N} (q_i -  t_i)
  = \frac{1}{N} \sum\limits_{i=1}^{N} q_i \times \Bigl(1 -
  \frac{\sum\limits_{\bm{c}_{\partial i}} \Theta\bigl(\sum\limits_{j\in\partial i} c_j
    -\theta_i \bigr) \prod\limits_{j\in\partial i} \bigl( \delta_{c_j}^0
    q_{j\rightarrow i}^{0,1} + \delta_{c_j}^1 t_{j\rightarrow i}^{1, 1} \bigr)}{
    \sum\limits_{\bm{c}_{\partial i}} \Theta\bigl(\sum\limits_{j\in\partial i} c_j
    - \theta_i \bigr) \prod\limits_{j\in\partial i} q_{j\rightarrow i}^{c_j,1}  
  } \Bigr) \; .
  \label{eq:phi}
\end{equation}

For the special case of RR networks of degree $D$ and uniform threshold $K$, we may assume that all the cavity quantities $t_{j\rightarrow i}^{1,1}$ are equal to the same value $t^{1,1}$, and then Eq.~(\ref{eq:sji11}) is simplified to
\begin{equation}
  \label{eq:sji11simple}
  t^{1,1} = \frac{1}{z} e^{-\beta} \Bigl[ \sum\limits_{d\geq K}^{D-1} C_{D-1}^d
    (q^{1,1})^d (q^{0,1})^{D-1-d} + C_{D-1}^{K-1} (t^{1,1})^{K-1} (q^{0,1})^{D-K}
    \Bigr] \; ,
\end{equation}
where $z$ is determined according to Eq.~(\ref{eq:zval}). Equation (\ref{eq:sji11simple}) can be solved numerically at a fixed point of the BP equation (\ref{eq:BPsimple}). The mean fraction $\phi$ of break nodes is then
\begin{equation}
  \label{eq:phirr}
  \phi = \rho \times \biggl(
  1 - \frac{\sum\limits_{d=K}^{D} C_{D}^{d} (t^{1,1})^d
    (q^{0,1})^{D-d}}{\sum\limits_{d=K}^D C_{D}^d (q^{1,1})^d
    (q^{0,1})^{D-d}} \biggr) \; .
\end{equation}

Notice that $t^{1,1}=q^{1,1}$ is always a root of Eq.~(\ref{eq:sji11simple}), and for this root we have $\phi = 0$ (i.e., there are no break nodes in an alliance configuration). It is easy to check that if the condition
\begin{equation}
  \label{eq:phicondition}
  \frac{1}{z} e^{-\beta} (K-1) C_{D-1}^{K-1} (q^{1,1})^{K-2} (q^{0,1})^{D-K}  > 1
\end{equation}
is \emph{not} satisfied, $t^{1,1} = q^{1,1}$ is the only root of Eq.~(\ref{eq:sji11simple}). However, if the condition (\ref{eq:sji11simple}) is satisfied, this root $t^{1,1}=q^{1,1}$ is no longer stable, and a stable root with $t^{1,1} < q^{1,1}$ can be obtained for Eq.~(\ref{eq:sji11simple}), for which $\phi > 0$ (i.e., a finite fraction of the nodes are break nodes).

Let us note that Eq.~(\ref{eq:phicondition}) will never be satisfied if $K = 1$ or $K=2$, which indicates that $\phi$ is always identical to zero for $K\leq 2$. Therefore, the threshold $K$ must at least be $3$ for a single-node induced collapse of an extensive alliance to occur. This conclusion can be further generalized, see more detailed discussions in Sec.~\ref{sec:ncd}.

\subsection{Damage propagation with local recruitment}

We now consider the damage propagation process with local recruitments. In this modified dynamics, if an originally active node (say $j$) has now only $\theta_j\! -\! 1$ active nearest neighbors but it has one or more persuadable inactive nearest neighbors, a randomly chosen persuadable inactive nearest neighbor (say node $k$) is recruited to the alliance (that is, $c_k$ flips from $0$ to $1$) to protect node $j$ from dropping to the inactive state.

To derive the corresponding mean field message-passing equations, let us consider once again the cavity factor graph in the lower-right panel of Fig.~\ref{fig:fg}. For the variable-node $(j, i)$ which is constrained only by the factor-node $j$,  we can compute through the following expression the probability $u_{j\rightarrow i}^{0, 1}$ that (1) this variable-node is staying in the composite state $(c_j, c_i) = (0, 1)$ and (2) node $j$ has less than $\theta_j$ active nearest neighbors:
\begin{equation}
  u_{j\rightarrow i}^{0,1} =
  \frac{1}{z_{j\rightarrow i}} \sum\limits_{\bm{c}_{\partial j\backslash i}}
  \Theta\bigl( \theta_j - 2 - \sum\limits_{k\in \partial j\backslash i} c_k \bigr)
  \prod_{k\in \partial j\backslash i} q_{k\rightarrow j}^{c_k, 0}
  \; ,
  \label{eq:aji01}
\end{equation}
where $z_{j\rightarrow i}$ is the same normalization constant as expressed in Eq.~(\ref{eq:zjtoi}). Considering the stabilizing effect of the local recruitment actions, the self-consistent equation (\ref{eq:sji11}) for the cavity probability distribution $t_{j\rightarrow i}^{1,1}$ is modified to
\begin{eqnarray}
  t_{j\rightarrow i}^{1,1} & = & \frac{1}{z_{j\rightarrow i}} e^{-\beta}
  \sum\limits_{\bm{c}_{\partial j\backslash i}} \biggl[
    \Theta\bigl(\sum\limits_{k\in\partial j\backslash i} c_k - \theta_j \bigr)
    \prod\limits_{k\in \partial j\backslash i} q_{k\rightarrow j}^{c_k,1}
    \nonumber \\
    & & + \delta_{\sum_{k\in \partial j\backslash i} c_k}^{\theta_j -1}
    \Bigl( \prod\limits_{k\in \partial j\backslash i} \bigl( \delta_{c_k}^0
    q_{k\rightarrow j}^{0, 1} + \delta_{c_k}^1 q_{k\rightarrow j}^{1,1} \bigr)
    - \prod\limits_{k\in \partial j\backslash i} \bigl( \delta_{c_k}^0
    u_{k\rightarrow j}^{0, 1} + \delta_{c_k}^1 q_{k\rightarrow j}^{1, 1} \bigr) \Bigr)
    \nonumber \\
    & & + \delta_{\sum_{k\in \partial j\backslash i} c_k}^{\theta_j -1}
    \prod\limits_{k\in \partial j\backslash i} \bigl(
    \delta_{c_k}^0 u_{k\rightarrow j}^{0, 1} + \delta_{c_k}^1 t_{k\rightarrow j}^{1, 1}
    \bigr) \biggr] \; .
  \label{eq:sji11modify}
\end{eqnarray}
There are three contributing terms within the square brackets of the above expression. The first term corresponds to the situation of node $j$ having more than $\theta_j$ active nearest neighbors; the second term corresponds to the situation of node $j$ having exactly $\theta_j$ active nearest neighbors and at the same time having at least one persuadable inactive nearest neighbors; the third term corresponds to the situation of node $j$ having no persuadable inactive nearest neighbor and having exactly $\theta_j$ ordinary active nearest neighbors.

Equation (\ref{eq:sji11modify}) can be rewritten in a slightly more compact form:
\begin{eqnarray}
  \hspace*{-0.5cm}
  t_{j\rightarrow i}^{1,1} & =  &  \frac{1}{z_{j\rightarrow i}} e^{-\beta}
  \sum\limits_{\bm{c}_{\partial j\backslash i}} \biggl[
    \Theta\bigl(\sum\limits_{k\in\partial j\backslash i} c_k +1 -\theta_j \bigr)
    \prod\limits_{k\in \partial j\backslash i} q_{k\rightarrow j}^{c_k,1}
    \nonumber \\
    & & - \delta_{\sum_{k\in \partial j\backslash i} c_k}^{\theta_j -1}
    \Bigl( \prod\limits_{k\in \partial j\backslash i} \bigl( \delta_{c_k}^0
    u_{k\rightarrow j}^{0, 1} + \delta_{c_k}^1 q_{k\rightarrow j}^{1, 1} \bigr) -
    \prod\limits_{k\in \partial j\backslash i} \bigl( \delta_{c_k}^0
    u_{k\rightarrow j}^{0, 1} + \delta_{c_k}^1 t_{k\rightarrow j}^{1,1} \bigr)
    \Bigr)
    \biggr] \; .
  \label{eq:sji11modify2}
\end{eqnarray}
In the special case of RR networks of degree $D$ and uniform threshold $K$, by assuming that $u_{j\rightarrow i}^{0,1} = u^{0, 1}$ and $t_{j\rightarrow i}^{1,1} = t^{1,1}$, we obtain from Eqs.~(\ref{eq:aji01}) and (\ref{eq:sji11modify2}) the following simplified equations
\begin{subequations}
  \begin{align}
    u^{0,1} & = \frac{1}{z} \sum\limits_{d=0}^{K-2} C_{D-1}^d (q^{1,0})^d
    (q^{0, 0})^{D-1-d} \; , \\
    t^{1,1} & = \frac{1}{z} e^{-\beta} \biggl[ \sum\limits_{d\geq K-1} C_{D-1}^d
      (q^{1,1})^d (q^{0,1})^{D-1-d} - C_{D-1}^{K-1} \bigl( (q^{1,1})^{K-1} -
      (t^{1,1})^{K-1} \bigr) (u^{0,1})^{D-K} \biggr]
    \; ,
    \label{eq:s11modsim}
  \end{align}
\end{subequations}
where $z$ is computed through Eq.~(\ref{eq:zval}). 

The mean fraction of break nodes with respect to the modified damage propagation process is denoted as $\psi$. It has the same expression as Eq.~(\ref{eq:phi}) for $\phi$, but with $t_{j\rightarrow i}^{1,1}$ being computed through Eq.~(\ref{eq:sji11modify2}). For the special case of RR networks of degree $D$ and uniform threshold $K$, the expression of $\psi$ is computed through the simplified equation (\ref{eq:phirr}), but with $t^{1,1}$ being computed through Eq.~(\ref{eq:s11modsim}). 

\subsection{Prerequisite for the phenomenon of sudden collapse}
\label{sec:ncd}

A necessary (but not sufficient) condition for the phenomenon of sudden collapse is that some threshold parameters $\theta_i$ should be greater than two. In other words, if $\theta_i \leq 2$ for all the nodes $i$ of the network, then every active node of an alliance configuration $\bm{c}$ must be ordinarily active and hence the breakdown of a single active node will never induce a global collapse of the alliance $A(\bm{c})$.

The reason for the above-mentioned prerequisite is easy to explain. An active node $i$ of the alliance configuration $\bm{c}$ is marginally stable if it has exactly $\theta_i$ active nearest neighbors. When one of these $\theta_i$ active nearest neighbors (say $j$) changes to be inactive, node $i$ will be pulled to the state $c_i = 0$ if it is not protected by any persuadable inactive nearest neighbor. If $\theta_i =1$ this induced breakdown of node $i$ will not induce new breakdowns; if $\theta_i = 2$, the induced breakdown of $i$ may pull the remaining active nearest neighbor (say $k$) to the inactive state if node $k$ was marginally stable in $\bm{c}$. But if the value of $\theta_k$ is also less than three, the induced breakdown of node $k$ will pull at most one nearest neighbor to the inactive state. Therefore, we see that if all the threshold parameters are less than three, the damage propagation process induced by the forced breakdown of an active node will extend as linear chains and will never branch out. Consequently the alliance $A(\bm{c})$ will not collapse but only mildly shrink.

We now complement this intuitive argument by theoretical analysis on the mean field equations of Sec.~\ref{sec:mftnr}. Let us define a quantity $\varepsilon_{j\rightarrow i}$ as
\begin{equation}
  \varepsilon_{j\rightarrow i} \equiv \frac{t_{j\rightarrow i}^{1,1}
    - q_{j\rightarrow i}^{1,1}}{q_{j\rightarrow i}^{1,1}} \; ,
\end{equation}
which quantify the relative difference between $t_{j\rightarrow i}^{1,1}$ and $q_{j\rightarrow i}^{1,1}$. If all the values $\varepsilon_{j\rightarrow i}$ are identical to zero, then the fraction of break nodes must be $\phi = 0$. Consider a node $j$ in a generic network. If $\theta_j=1$, it is easy to verify from Eq.~(\ref{eq:sji11}) that $t_{j\rightarrow i}^{1,1} = q_{j\rightarrow i}^{1,1}$, i.e., $\varepsilon_{j\rightarrow i} = 0$ for every $i\in \partial j$. If the threshold $\theta_j = 2$ for node $j$, we obtain from Eq.~(\ref{eq:sji11}) and Eq.~(\ref{eq:BP:d}) that
\begin{equation}
  \varepsilon_{j\rightarrow i} = \sum\limits_{k\in \partial j\backslash i}
             W^{j\rightarrow i}_{k\rightarrow j} \varepsilon_{k\rightarrow j} \; ,
             \label{eq:epsilonite}
\end{equation}
where the matrix element $W^{j\rightarrow i}_{k\rightarrow j}$ is expressed as
\begin{equation}
  W^{j\rightarrow i}_{k\rightarrow j} =
  \frac{q_{k\rightarrow j}^{1,1} \prod\limits_{l\in \partial j\backslash i,k}
    q_{l\rightarrow j}^{0,1}}{
    \sum\limits_{k^\prime \in \partial j\backslash i} q_{k^\prime \rightarrow j}^{1,1}
    \prod\limits_{l\in \partial j\backslash i, k^\prime} q_{l\rightarrow j}^{0,1}
    + \sum\limits_{\bm{c}_{\partial j\backslash i}} \Theta\bigl(
    \sum\limits_{l\in \partial j\backslash i} c_l -2) \prod\limits_{l\in \partial j\backslash i} q_{l\rightarrow j}^{c_l, 1}
  } \; .
\end{equation}
Notice that all the matrix elements $W^{j\rightarrow i}_{k\rightarrow j}$ are non-negative, and that
\begin{equation}
  \sum\limits_{k\in \partial j\backslash i} W^{j\rightarrow i}_{k\rightarrow j} \leq 1
  \; .
\end{equation}
Therefore the maximum eigenvalue $\lambda_{max}$ of the matrix formed by all the elements $W^{j\rightarrow i}_{k\rightarrow j}$ of the network is less than unity. As a consequence, the iterative equation (\ref{eq:epsilonite}) has only a fixed point of $\varepsilon_{j\rightarrow i} = 0$ (for all nodes $j$ and all  $i\in \partial j$), and then the fraction $\phi$ of break nodes in the alliance configurations $\bm{c}$ must be identical to zero.

\clearpage

\section{Mean field theory on kinetic alliance configurations}

A kinetic alliance configuration is the result of an irreversible pruning process, starting from an initial activity pattern in which each node $i$ is active ($c_i \! = \! 1$) with probability $p$ and inactive ($c_i\! = \! 0$) with probability $1\! -\! p$. An initially active node (say $i$) will decay to be inactive during this pruning process if it has less than $\theta_i$ active nearest neighbors. If an initially active node (say $j$) remains to be active after the whole pruning process, it must have $\theta_j$ or more active nearest neighbors in the final alliance configuration $\bm{c}$. The mean field theory for equilibrium alliances (Sec.~\ref{sec:mfea}) can readily be adapted for the case of kinetic alliances, and because there is no need to consider the Boltzmann weight of each alliance configuration, the mean field equations are much simplified.

\subsection{Mean fraction of active nodes}

Consider a node $i$ of a random network $G$. The probability $q_i$ of this node being active in a kinetic alliance is expressed as
\begin{equation}
  \label{eq:kinetic_qi}
  q_i = p \sum\limits_{\bm{c}_{\partial i}}
  \Theta\bigl( \sum_{j\in \partial i} c_j -\theta_i \bigr)
  \prod\limits_{j\in \partial i} \bigl(
  \delta_{c_j}^0 (1 - \alpha_{j\rightarrow i}) +
  \delta_{c_j}^1 \alpha_{j\rightarrow i} \bigr) \; .
\end{equation}
In this expression, $\alpha_{j\rightarrow i}$ is the probability of a nearest neighbor $j$ being active in a kinetic alliance if node $i$ is always active. The self-consistent equation for this cavity probability is
\begin{equation}
  \label{eq:alphaji}
  \alpha_{j\rightarrow i}  = p
  \sum\limits_{\bm{c}_{\partial j\backslash i}}
  \Theta\bigl( \sum_{k\in \partial j\backslash i} c_k + 1 -\theta_j \bigr)
  \prod\limits_{k\in \partial j\backslash i} \bigl( \delta_{c_k}^0 (1-
  \alpha_{k\rightarrow j}) + \delta_{c_k}^1 \alpha_{k\rightarrow j} \bigr) \; .
\end{equation}
The mean fraction $\rho$ of active nodes in the kinetic alliance configurations is then
\begin{equation}
  \rho = \frac{1}{N} \sum\limits_{i=1}^{N} q_i \; .
\end{equation}

Notice that if all the cavity probabilities $\alpha_{j\rightarrow i}$ are equal to zero, Eq.~(\ref{eq:alphaji}) is surely satisfied and the corresponding value of $\rho$ is simply $\rho \! = \! 0$. When $p$ is larger than certain critical value $p_c$ Eq.~(\ref{eq:alphaji}) may have a non-trivial fixed point $\alpha_{j\rightarrow i} \! = \! \alpha_{j\rightarrow i}^* > 0$. We now derive an equation for determining the critical value $p_c$ as follows. Consider a small perturbation $\varepsilon_{j\rightarrow i}$ to each cavity probability $\alpha_{j\rightarrow i}$, such that
\begin{equation}
  \alpha_{j\rightarrow i} =  \alpha_{j\rightarrow i}^* + \varepsilon_{j\rightarrow i}
  \; .
\end{equation}
Then according to Eq.~(\ref{eq:alphaji}) the evolution of the perturbation $\varepsilon_{j\rightarrow i}$ is
\begin{equation}
  \label{eq:kineticepsilon}
  \varepsilon_{j\rightarrow i} \ \Leftarrow \
  p \sum\limits_{k\in \partial j\backslash i}
  A_{j\rightarrow i}^{k\rightarrow j} \varepsilon_{k\rightarrow j} \; 
\end{equation}
to linear order of $\varepsilon_{k\rightarrow j}$ values. Here the matrix element $A_{j\rightarrow i}^{k\rightarrow j}$ is defined by
\begin{equation}
  A_{j\rightarrow i}^{k\rightarrow j} \equiv
  \sum\limits_{\bm{c}_{\partial j\backslash i,k}}
  \delta_{\sum_{l\in \partial j\backslash i, k} c_l}^{\theta_j - 2}
    \prod\limits_{l\in \partial j\backslash i, k} \bigl(
    \delta_{c_l}^0 (1-\alpha_{l\rightarrow j}^*) +
    \delta_{c_l}^1 \alpha_{l\rightarrow j}^*\bigr)
    \; ,
\end{equation}
with $\partial j\backslash i, k$ denoting the subset of $\partial j$ after removing nodes $i$ and $k$, and $\bm{c}_{\partial j\backslash i, k} \! \equiv \! \{c_l : l \in \partial j\backslash i, k \}$. Let us denote by $\lambda_{max}$ the maximum eigenvalue of this matrix. From Eq.~(\ref{eq:kineticepsilon}) we see that the nontrivial fixed point will be stable if (and only if)
\begin{equation}
 p \, |\lambda_{max} | \ < \ 1 \; .
\end{equation}
Therefore the critical value $p_c$ corresponds to the point that $ p\, |\lambda_{max}|   = 1$.

\begin{figure}[t]
  \centering
  \subfigure[]{
    \includegraphics[angle=270,width=0.4\textwidth]{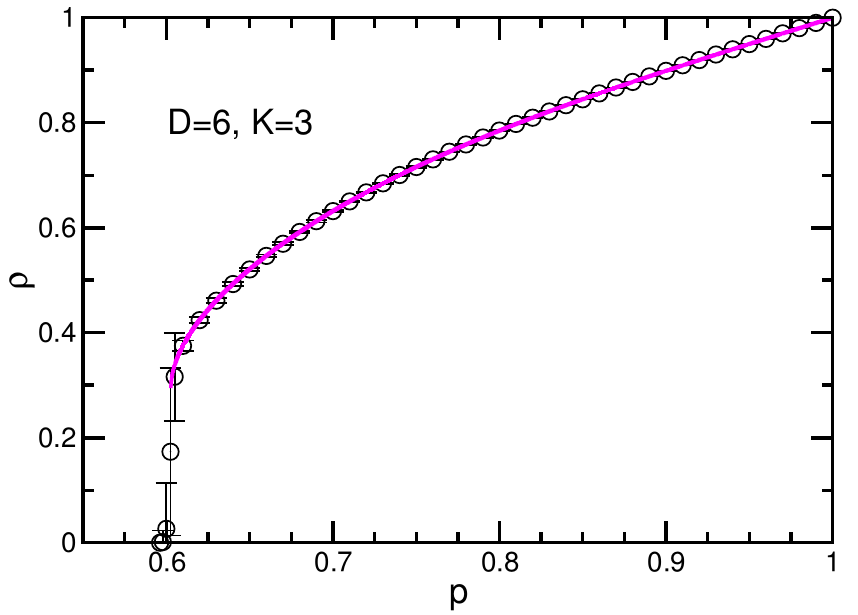}
  }
  \subfigure[]{
    \includegraphics[angle=270,width=0.4\textwidth]{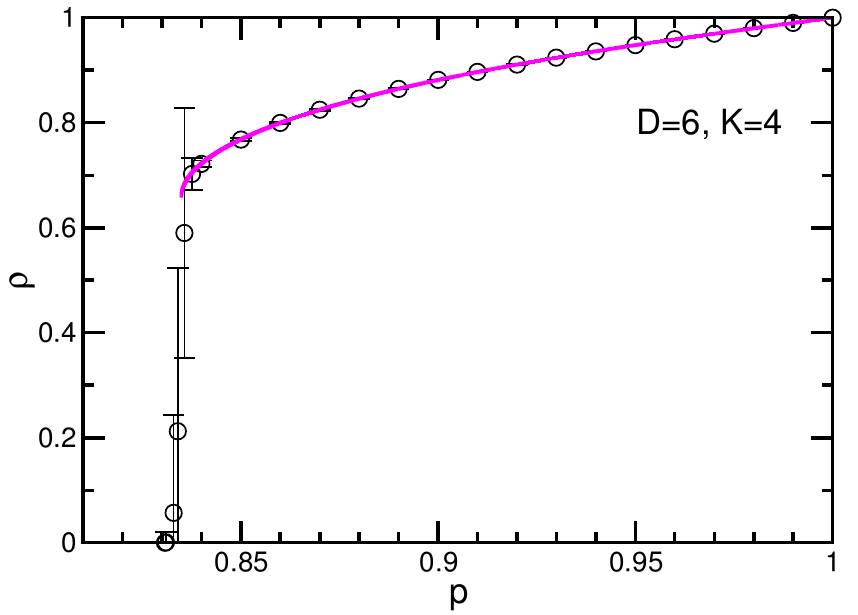}
  }

  \subfigure[]{
    \includegraphics[angle=270,width=0.4\textwidth]{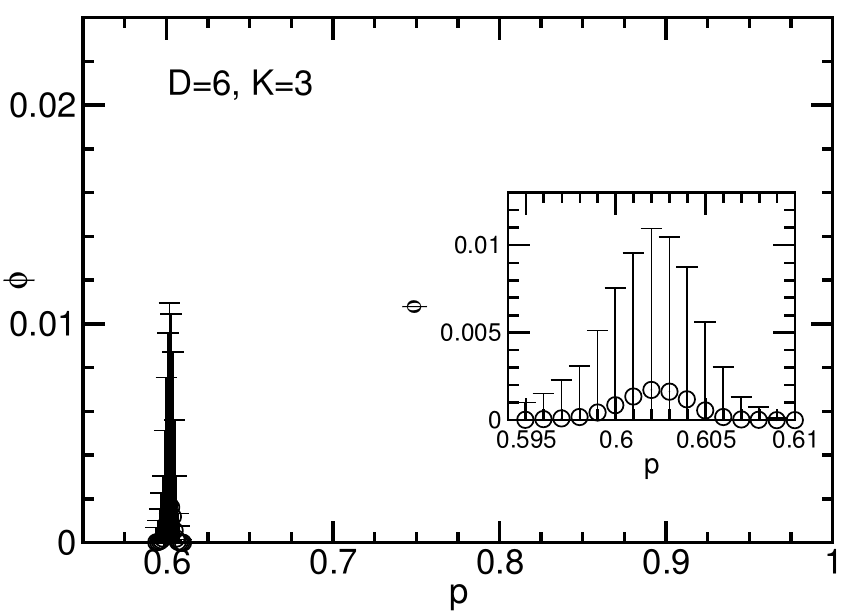}
  }
  \subfigure[]{
    \includegraphics[angle=270,width=0.4\textwidth]{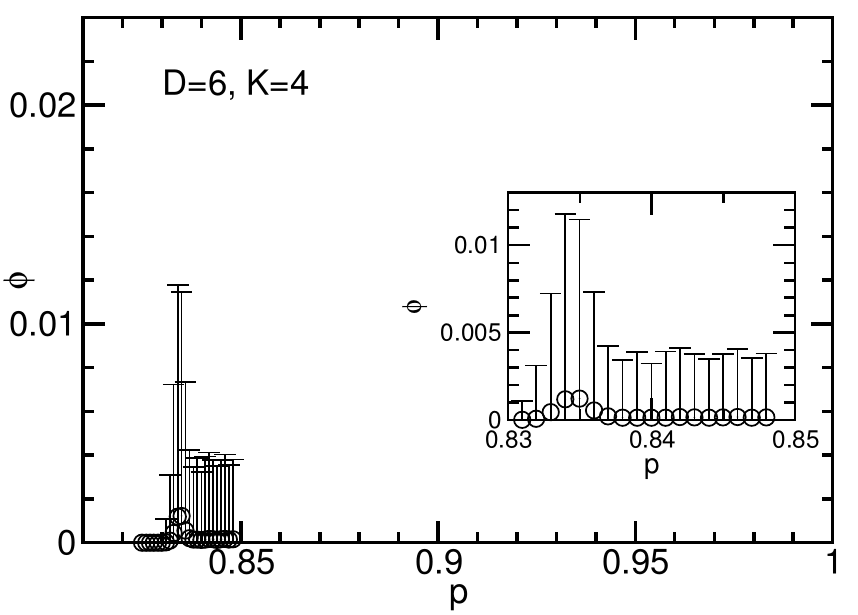}
  }
  \caption{
    \label{fig:KineticRRD6}
    Kinetic alliance configurations in regular random network of node degree $D\! = \! 6$, assuming uniform threshold value $K\! = \! 3$ (a) and $K\! = \! 4$ (b). Solid lines are theoretical results obtained through Eqs.~(\ref{eq:rhorrk}) and (\ref{eq:alpharrk}); circles with error bars are results obtained by running the irreversible pruning process on $2300$ independent initial activity patterns of a single RR network (size $N\! = \! 32768$). The parameter $p$ is the initial fraction of active nodes, and $\rho$ is the final fraction of active nodes.
  }
\end{figure}

In the case of RR networks which have uniform node degree $D$, if in addition all the thresholds $\theta_i$ are the same ($\theta_i \! =\! K$), we may assume that $\alpha_{j\rightarrow i}\! = \! \alpha$ for all the edges of the network. Then Eq.~(\ref{eq:alphaji}) becomes
\begin{equation}
  \label{eq:alpharrk}
  \alpha = p \sum\limits_{m=K-1}^{D-1} C_{D-1}^m \alpha^{m} (1-\alpha)^{D-1-m} \; .
\end{equation}
When $p$ is larger than the critical value $p_c$ this equation has a positive stable root, and the corresponding mean fraction of active nodes is
\begin{equation}
  \label{eq:rhorrk}
  \rho = p \sum\limits_{m=K}^{D} C_{D}^{m} \alpha^{m} (1-\alpha)^{D-m} \; .
\end{equation}
Some of the theoretical predictions obtained for $D\! = \! 6$ are shown in Fig.~\ref{fig:KineticRRD6}, which are in good agreement with computer simulation results.  At $K\! = \! 3$ we see that $p_{\rm{c}} \! = \! 0.6028$, which corresponds to the minimum active-node density $\rho_{\rm{k}}\! = \! 0.2942$; at $K\! = \! 4$ the critical value is $p_{\rm{c}} \! = \! 0.8349$ and the relative size of the minimum kinetic alliances is $\rho_{\rm{k}} \! = \! 0.6574$. Notice that the minimum value $\rho_{\rm{k}}$ achieved by the irreversible pruning process is much higher than the minimum relative size of equilibrium alliances.

\subsection{Damage cascading in kinetic alliances}

We expect that kinetic alliance configurations, as the result of an irreversible pruning process, should be quite robust again single-node perturbations. The intuitive reason behind this expectation is quite simple: if instead such an alliance configuration $\bm{c}$ contains an extensive number of break nodes, it should be extremely unlikely for all these break nodes to be active in the initial activity pattern. Our numerical simulation results indeed confirm that the number of break nodes becomes non-zero only when the value of $p$ is approaching $p_{\rm{c}}$ very closely.

The robustness of kinetic alliance configurations to single-node perturbations can also be anticipated from the mean field theory. To compute the mean fraction of break nodes in kinetic alliance configurations, we notice that the probability $t_i$ of node $i$ being active but not being a break node is
\begin{equation}
  t_i =  \sum\limits_{\bm{c}_{\partial i}}
  \Theta\bigl( \sum_{j\in \partial i} c_j -\theta_i \bigr)
  \prod\limits_{j\in \partial i} \bigl(
  \delta_{c_j}^0 (1 - \alpha_{j\rightarrow i}) +
  \delta_{c_j}^1 \gamma_{j\rightarrow i} \bigr) \; ,
\end{equation}
where $\gamma_{j\rightarrow i}$ is the probability of a nearest neighboring node $j$ (1) being active if node $i$ always keep to be active and (2) if node $i$ flips to be inactive the damage cascade relayed by the link $(i, j)$ will only cause a tree-formed (small) avalanche. The self-consistent expression for this cavity probability is
\begin{eqnarray}
  \gamma_{j\rightarrow i} & = & p \sum\limits_{\bm{c}_{\partial j\backslash i}}
  \Bigl[ \Theta\bigl( \sum_{k\in \partial j\backslash i} c_k -\theta_j \bigr)
    \prod\limits_{k\in \partial j\backslash i} \bigl( \delta_{c_k}^0 (1-
    \alpha_{k\rightarrow j}) + \delta_{c_k}^1 \alpha_{k\rightarrow j} \bigr)
    \nonumber \\
    & & \quad \quad + 
    \delta_{\sum_{k\in \partial j\backslash i} c_k}^{\theta_j -1}
    \prod\limits_{k\in \partial j\backslash i} \bigl( \delta_{c_k}^0 (1-
    \alpha_{k\rightarrow j}) + \delta_{c_k}^1 \gamma_{k\rightarrow j} \bigr)
    \Bigr] \; .
  \label{eq:gammaji}
\end{eqnarray}

\begin{figure}[t]
  \centering
  \subfigure[]{
    \label{fig:kineticslopeRR}
    \includegraphics[angle=270,width=0.4\textwidth]{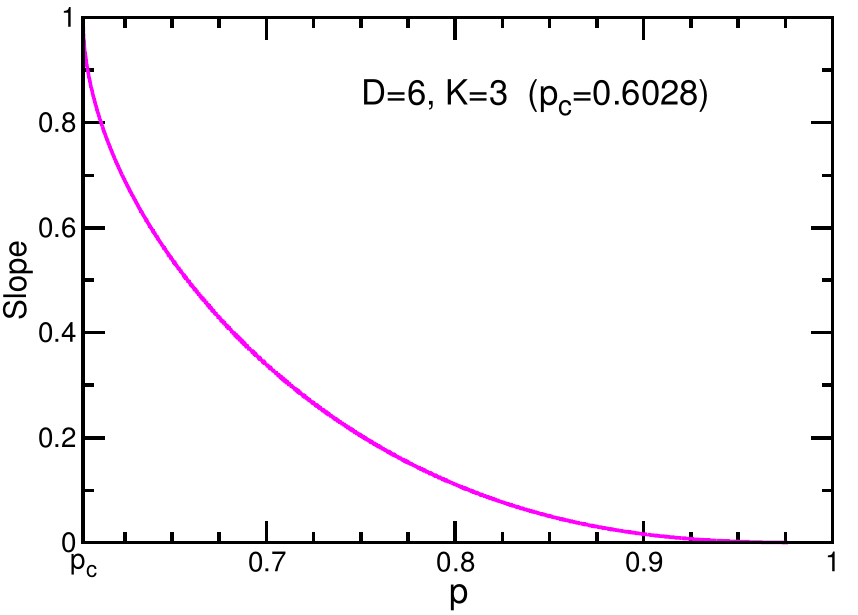}
  }
  \subfigure[]{
        \label{fig:kineticslopeER}
    \includegraphics[angle=270,width=0.4\textwidth]{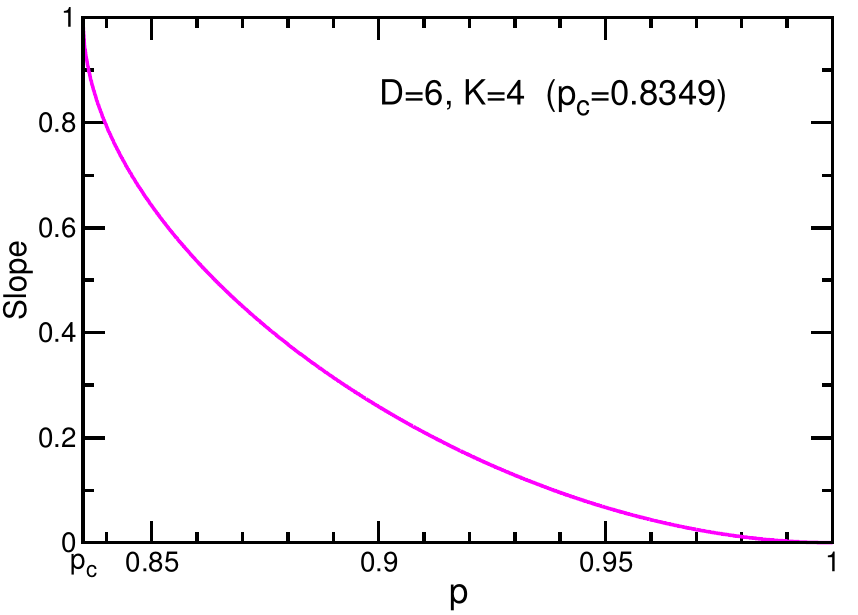}
  }
  \caption{
    \label{fig:KineticRRD6slope}
    The slope $\frac{\rm{d} g(\gamma)}{\rm{d} \gamma}$ as defined by Eq.~(\ref{eq:dgdg}) for the RR network ensemble of degree $D\! = \! 6$ and uniform threshold $K\! = \! 3$ (a) and $K\! = \! 4$ (b). This slope approaches unity as the initial fraction $p$ of active nodes is decreased to the critical value $p_{\rm{c}}$.
  }
\end{figure}

In the case of RR networks of degree $D$ with uniform threshold $K$, the cavity probability $\gamma_{j\rightarrow i}$ is the same value (denoted as $\gamma$) for all the links, so Eq.~(\ref{eq:gammaji}) is simplified to
\begin{equation}
  \label{eq:gRR}
  \gamma = g(\gamma) \; ,
\end{equation}
where the function $g(\gamma)$ is defined as
\begin{equation}
  g(\gamma) = \alpha - p C_{D-1}^{K-1} (1-\alpha)^{D-K} \bigl(\alpha^{K-1} -
  \gamma^{K-1} \bigr) 
  \; .
\end{equation}
Notice that $g(\gamma)$ is a convex increasing function of $\gamma$ for $\gamma \! \in \! [0, \alpha]$, and that $g(\alpha)\! =\! \alpha$, so Eq.~(\ref{eq:gRR}) has a trivial fixed point $\gamma\! =\! \alpha$. The slope of $g(\gamma)$ at $\gamma \! = \! \alpha$ is
\begin{equation}
  \label{eq:dgdg}
  \frac{{\rm d} g(\gamma)}{{\rm d} \gamma} \biggr|_{\gamma = \alpha}
  = p (K-1) C_{D-1}^{K-1} (1-\alpha)^{D-K} \alpha^{K-2}
  \; .
\end{equation}
It turns out that this slope is always below unity for $p \! > \! p_{\rm{c}}$ and it is equal to unity exactly at $p_{\rm{c}}$ (Fig.~\ref{fig:KineticRRD6slope}). This means that $\gamma \! = \! \alpha$ is the only solution of Eq.~(\ref{eq:gRR}) for the whole region of $p\! \in \! [p_{\rm{c}}, 1]$, and consequently the fraction of bread nodes in the kinetic alliance configurations is zero. At $p_{\rm{c}}$ this fixed point is marginally stable, and break points start to emerge and the kinetic alliance configurations then disappear.

For more general networks, let us assume $\gamma_{k\rightarrow j}$ in Eq.~(\ref{eq:gammaji}) deviates slightly from the fixed-point value $\alpha_{k\rightarrow j}^*$, that is,
\begin{equation}
  \gamma_{k\rightarrow j} = \alpha_{k\rightarrow j}^* + \varepsilon_{k\rightarrow j}
  \; .
\end{equation}
Then according to Eq.~(\ref{eq:gammaji}), to linear order in $\varepsilon_{k\rightarrow j}$, the perturbation $\varepsilon_{j\rightarrow i}$ to $\gamma_{j\rightarrow i}$ evolves according to
\begin{equation}
  \varepsilon_{j\rightarrow j} \ \Leftarrow \
  p \sum\limits_{k\rightarrow j\backslash i} A_{j\rightarrow i}^{k\rightarrow j}
  \varepsilon_{k\rightarrow j}
  \; ,
\end{equation}
which is identical in form to Eq.~(\ref{eq:kineticepsilon}). This means that if $\alpha_{j\rightarrow i}^*$ is a stable fixed point of Eq.~(\ref{eq:alphaji}), i.e. $\rho \! > \! 0$, then $\gamma_{j\rightarrow i} \! = \! \alpha_{j\rightarrow i}^*$ is also a stable fixed point of Eq.~(\ref{eq:gammaji}), namely $\phi \! = \! 0$ (the number of break nodes in the surviving kinetic alliance configuration is not extensive).

\clearpage

\section{Microcanonical Monte Carlo simulation}

We perform microcanonical Monte Carlo (MMC) simulation to sample a set of alliance configurations $\bm{c}$, all of which having the same density $\rho$ of active nodes~\cite{Creutz-1983}. First, an objective value $E_{\rm{o}}$ of active nodes is set to be $E_{\rm{o}} \equiv N \rho$, where $N$ is the network size. Starting from the initial time $t=0$ and an initial alliance configuration $\bm{c}(0)$ whose number of active nodes is less than or equal to $E_{\rm{o}}$,  a new alliance configuration $\bm{c}^\prime$ is proposed for the network at each elementary time step $\Delta t = \frac{1}{N}$ of the MMC dynamics. If the number $E^\prime$ of active nodes in $\bm{c}^\prime$ is less than or equal to $E_{\rm{o}}$, then $\bm{c}^\prime$ is accepted as the configuration of the network at time $t+\Delta t$, that is, $\bm{c}(t+\Delta t) = \bm{c}^\prime$. If $E^\prime > E_{\rm{o}}$, however, the proposed configuration $\bm{c}^\prime$ is ignored and the configuration of the network is unchanged, $\bm{c}(t+\Delta t) = \bm{c}(t)$. At each value of $\rho$, we typically sample with equal statistical weight $\mathcal{N} = 10^5$ alliance configurations at unit time interval (one unit time corresponds to $N$ configuration transition trials) to evaluate the order parameters $\phi$ and $\psi$.

The simulation results shown in Fig.~2(b) and Fig.~2(c) of the main text and in Fig.~\ref{fig:ERD6k4} and Fig.~\ref{fig:All4cores} are obtained by decreasing the energy density $\rho$ slowly from a high value. The characteristic equlibrium time of the MMC dynamics becomes more and more longer as the value of $\rho$ decreases, so we can not reach the global minimum energy density $\rho_0$ by this method.

Another way of preparing an initial alliance configuration $\bm{c}(0)$ at a given energy density $\rho$ is through the Clamp-Alliance message-passing algorithm of Ref.~\cite{Xu-etal-2018}. Starting from $\bm{c}(0)$ we then run the MMC dynamics a long time to drive the system towards equilibrium. The advantage of this method is that we can cover energy density values $\rho$ down to the vicinity of the global minimum value $\rho_0$.  The simulation results shown in Fig.~4(b) and Fig.~4(d) of the main text are obtained using such initial conditions.

Extending the algorithm of Ref.~\cite{Xu-etal-2018}, here we employ single-node flips and tree flips to propose a new alliance configuration $\bm{c}^\prime$ based on the current configuration $\bm{c}$ of the network. Special care is taken to guarantee detailed balance, namely the transition $\bm{c} \rightarrow \bm{c}^\prime$ and the reverse transition $\bm{c}^\prime \rightarrow \bm{c}$ are equally likely to occur in the MMC evolution process. In our MMC algorithm the single-node flips and tree flips are tried with equal probability (i.e., one-half).

\subsection{Single-node state flip}

A single-node flipping trial is a proposed state change $c_i \rightarrow 1-c_i$ for a node $i$, with the new alliance configuration $\bm{c}^\prime$ differing from the old configuration $\bm{c}$ only at position $i$. Because $A(\bm{c}^\prime)$ needs to be a valid alliance, node $i$ must be chosen from one of two sets $V_{1\rightarrow 0}(\bm{c})$ and $V_{0\rightarrow 1}(\bm{c})$. Here $V_{1\rightarrow 0}(\bm{c})$ contains all the active nodes of $\bm{c}$ which can be flipped to the inactive state without affecting its active nearest neighbors, and $V_{0\rightarrow 1}(\bm{c})$ contains all the persuadable inactive nodes of $\bm{c}$~\cite{Xu-etal-2018}.  We conduct single-node flipping trials under the condition of detailed balance according to the following recipe:
\begin{enumerate}
\item[1.] Generate a uniform real random number $x_1$ in $[0, 1)$; if $x_1 < \frac{|V_{1\rightarrow 0}(\bm{c})|}{|V_{1\rightarrow 0}(\bm{c})|+|V_{0\rightarrow 1}(\bm{c})|}$, where $|V|$ denotes the cardinality of the node set $V$, randomly choose an active node $i$ from set $V_{1\rightarrow 0}(\bm{c})$ and propose a flip to $c_i=0$,  otherwise randomly choose an inactive node $j$ from $V_{0\rightarrow 1}(\bm{c})$ and propose a flip to $c_j=1$. 
\item[2.] If the number $E(\bm{c}^\prime)$ of active nodes in the resulting new configuration $\bm{c}^\prime$ does not exceed $E_{\rm{o}}$, then generate another uniform real random number $x_2$ in $[0, 1)$ and accept $\bm{c}^\prime$ as the new configuration of the network if $x_2 < A_{s}(\bm{c}\rightarrow \bm{c}^\prime)$. However, if $E(\bm{c}^\prime) > E_{\rm{o}}$ or $x_2 \geq A_{s}(\bm{c}\rightarrow \bm{c}^\prime)$, then discard $\bm{c}^\prime$ and let the network keep the old configuration $\bm{c}$. Here the acceptance rate $A_s(\bm{c}\rightarrow \bm{c}^\prime)$ is
  \begin{equation}
    A_s(\bm{c}\rightarrow \bm{c}^\prime) = \min\Bigl(1, \
    \frac{|V_{1\rightarrow 0}(\bm{c})| +  |V_{0\rightarrow 1}(\bm{c})|}
         {|V_{1\rightarrow 0}(\bm{c}^\prime)|+ |V_{0\rightarrow 1}(\bm{c}^\prime)|}
         \Bigr) \; .
  \end{equation}
\end{enumerate}

\subsection{Tree state flip}

Two or more nodes of the same state, which form a connected tree-formed subnetwork in the network, are flipped simultaneously through the tree flipping trials. These updating trials extend the simpler bridge (or chain) flipping trials employed in Ref.~\cite{Xu-etal-2018}. They propose longer-distance hops in the space of alliance configurations and help shortening the equilibrium time of the MMC evolution process.

An inactive node (say $i$) is regarded as a candidate root of a possible inactive tree if $i$ has exactly $\theta_i \!-\!1$ active nearest neighbors. When such a node $i$ is flipped to the state $c_i\!=\!1$ one of its inactive nearest neighbors must also be flipped. The set of all such inactive nodes of the alliance configuration $\bm{c}$ is denoted as $T_{0\rightarrow 1}(\bm{c})$. An active node $j$ is regarded as a candidate root of a possible active tree if (1) $j$ has exactly $\theta_j$ active nearest neighbors and (2) one and only one of these active nearest neighbors (say $k$) has exactly $\theta_k$ active nearest neighbors itself. When such a node $j$ is flipped to $c_j\!=\!0$ the active nearest neighbor $k$ must also be flipped. The set of all such active nodes $j$ in $\bm{c}$ is denoted as $T_{1\rightarrow 0}(\bm{c})$.

To construct an inactive tree $Y$ for the configuration $\bm{c}$ we proceed as follows:
\begin{enumerate}
\item[1.] Draw an inactive node from the set $T_{0\rightarrow 1}(\bm{c})$ uniformly at random and consider it as the only node at layer $l = 1$ of a nascent inactive tree.
\item[2.] For each and every node (say $i$) at the newly extended layer $l$ of the inactive tree, construct a node set $C_i$ by inserting to this set  all the inactive nearest neighbors (say $k$) of node $i$ which (1) are \emph{not} connected to any node in the earlier layers $l^\prime$ ($l^\prime < l$) of this tree and (2) have less than $\theta_k$ active nearest neighbors. Suppose node $i$ needs $m_i$ of these inactive nodes in $C_i$ to be flipped to make it be surrounded by exactly $\theta_i$ active nearest neighbors. If $m_i > 0$, then pick $m_i$ nodes from set $C_i$ uniformly at random and add them to the next layer $l+1$ of the inactive tree. The total number of possible ways of picking these $m_i$ neighbors is $\frac{ |C_i|  !}{m_i ! ( |C_i| - m_i)!}$, where $|C_i|$ is the cardinality of set $C_i$.
\item[3.] After no more nodes need to be added to the new layer $l+1$, then check if the so-far constructed subnetwork is really a tree. If it contains at least one loop then it is discarded; otherwise, repeat step (2) to further extend the inactive tree.
\end{enumerate}
If this tree construction is successful and we obtain the final inactive tree $Y$, then we asign it a surprisal scale $W_{Y}$ as
\begin{equation}
  W_{Y} = \prod\limits_{i\in Y} \frac{(|C_i|)!}{m_i ! ( |C_i| - m_i) !} \; .
\end{equation}

An active node $j$ is considered to be critical (or marginally stable) if it has exactly $\theta_j$ active nearest neighbors. To construct an active tree $Y$ of critical nodes we proceed as follows:
\begin{enumerate}
\item[1.] Draw an active node from the candidate set $T_{1\rightarrow 0}(\bm{c})$ uniformly at random and consider it as the only node at layer $l = 1$ of a nascent active tree.
\item[2.] For each and every node (say $j$) at the newly extended layer $l$ of the active tree, add all its critical active nearest neighbors (say $k$) to the next layer $(l+1)$ of the tree if these nodes $k$ do not belong to earlier layers of this tree.
\item[3.] Check if there is any loop in the resulting subgraph. If there is at least one loop, the subnetwork is discarded; otherwise, repeat step (2) to further extend the active tree.
\end{enumerate}

We conduct tree flipping trials under the condition of detailed balance according to the following recipe, which achieve a change of the alliance configuration $\bm{c}$ to a new configuration $\bm{c}^\prime$: 
\begin{enumerate}
\item[1.] With conditional probability $\frac{|T_{0\rightarrow 1}(\bm{c})|}{|T_{0\rightarrow 1}(\bm{c})| + |T_{1\rightarrow 0}(\bm{c})|}$,  a tree addition trial is performed: an inactive tree $Y$ is generated according to the above-mentioned protocol and, if it is a valid inactive tree and $E(\bm{c}^\prime) \leq E_{\rm{o}}$, the whole tree $Y$ is flipped and accepted with probability
  \begin{equation}
    \label{eq:Am01gen}
    A_{t}^{0\rightarrow 1}(\bm{c}\rightarrow \bm{c}^\prime) =
    \min\biggl(1, \ \frac{|T_{0\rightarrow 1}(\bm{c})|
      +|T_{1\rightarrow 0}(\bm{c})|}{|T_{0\rightarrow 1}(\bm{c}^\prime)|
      + |T_{1\rightarrow 0}(\bm{c}^\prime)|}
    W_{Y} \biggr) \; .
  \end{equation}
\item[2.] With the remaining conditional probability $\frac{|T_{1\rightarrow 0}(\bm{c})|}{|T_{0\rightarrow 1}(\bm{c})| + |T_{1\rightarrow 0}(\bm{c})|}$ a tree deletion trial is performed: an active tree $Y$ is generated according to the above-mentioned protocol and, if it is a valid active tree, the whole tree is flipped and accepted with probability
  \begin{equation}
    \label{eq:Am10gen}
    A_{t}^{1\rightarrow 0}(\bm{c}\rightarrow \bm{c}^\prime)
    = \min\biggl(1, \ \frac{|T_{0\rightarrow 1}(\bm{c})|+
      |T_{1\rightarrow 0}(\bm{c})|}{|T_{0\rightarrow 1}(\bm{c}^\prime)|+
      |T_{1\rightarrow 0}(\bm{c}^\prime)|}
    \frac{1}{W_{Y}} \biggr) \; ,
  \end{equation}
  where $W_{Y}$ is the surprisal scale of the resulting inactive tree $Y$ \emph{after} the flip.
\end{enumerate}

\clearpage

\section{Supplementary simulation and theoretical results}

\subsection{Random regular (RR) networks}

\begin{figure}[h!]
  \centering
  \includegraphics[angle=270,width=0.5\textwidth]{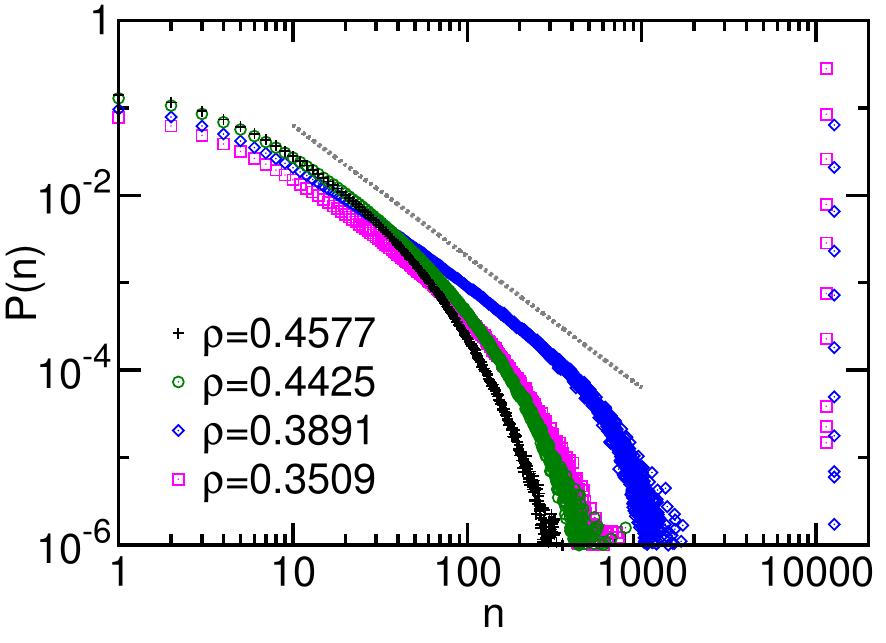}
  \caption{\label{fig:profileRRk3}
    Distribution $P(n)$ of avalanche size $n$ in single-node-induced damage cascading processes, obtained on a single RR network of size $N\! =\! 62584$ and degree $D\! =\! 6$, with uniform threshold $K\! =\! 3$. This distribution is obtained by examining $32000$ equilibrium alliance configurations at fixed energy density $\rho$, with $\rho$ being $0.4577$ (pluses), $0.4425$ (circles), $0.3891$ (diamonds), and $0.3509$ (squares). The thin dotted line mark the power-law behavior $P(n) \propto n^{-\frac{3}{2}}$.
  }
\end{figure}

We show the distribution $P(n)$ of damage sizes $n$ associated with the forced breakdown of a single active node (without local protections), obtained on a RR network of size $N=32768$ and node degree $D=6$ with uniform threshold $K\! = \! 3$ (Fig.~\ref{fig:profileRRk3}).

When the relative size $\rho$ of the alliances is much larger than the value $\rho_{\rm wt} = 0.3885$ of the weak tipping point (e.g., $\rho \! = \! 0.4425$ or $0.4577$), the distribution $P(n)$ decays quickly with $n$ and the avalanche size $n$ is much less than $N$. When $\rho$ approaches $\rho_{\rm wt}$ or becomes smaller than this critical value (e.g., $\rho\! = 0.3891$ or $0.3509$), the avalanche size $n$ sometimes is equal to the size of the whole alliance $A(\bm{c})$ and the distribution $P(n)$ becomes bimodal, signifying the existence of an extensive number of break nodes in the alliance configurations.

\begin{figure}[h!]
  \centering
  \includegraphics[angle=270,width=0.5\textwidth]{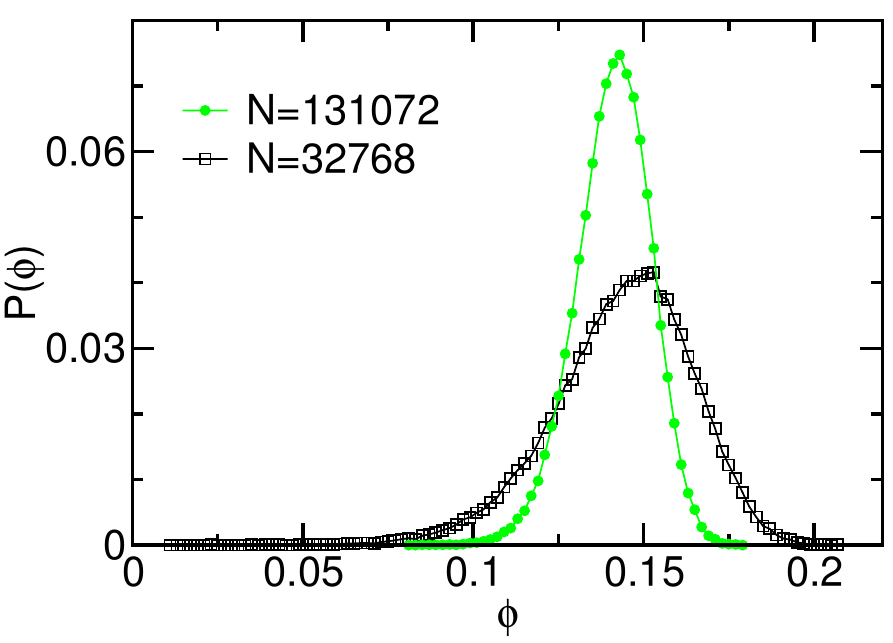}
  \caption{\label{fig:phidist}
    Probability distribution $P(\phi)$  of the fraction $\phi$ of break nodes in a single RR network of size $N \! = \! 32768$ (squares) or $N\! = \! 131072$ (circles), with degree $D\! = \! 6$ and uniform threshold $K\! = \! 3$. We sample a total number of $10^5$ alliance configurations for each network instance by the MMC algorithm at unit time interval, and compute the fraction $\phi$ of break nodes for each of them. The energy density is fixed at $\rho = 0.35$, below the critical value $\rho_{\rm wt} = 0.3885$ of the weak tipping point.
    }
\end{figure}

The simulation results reported in Fig.~2(b) and Fig.~2(c) of the main text show that the standard deviations of the order parameters $\phi$ and $\psi$ are relatively large, especially when $\rho$ is close to the weak tipping point $\rho_{\rm{wt}}$ (for $\phi$) or the strong tipping point $\rho_{\rm{st}}$ (for $\psi$). Such relatively large fluctuations are caused by finite-size effect and they should reduce with the system size $N$. For example, as shown in Fig.~\ref{fig:phidist}, when the network size $N$ increases the probability profile of $\phi$ becomes more concentrated, and then the standard deviation of $\phi$ is much reduced.

\subsection{Erd\"os-R\'enyi (ER) networks}

An ER network is generated by setting up $M$ links completely at random between $N$ nodes but prohibiting multiple links between the same pair of nodes and self-links from a node to itself. The mean node degree $D \equiv \frac{2 M}{N}$ of an ER network is a real value in general. When the network size $N$ is large, the number of nearest neighbors a node $i$ has, its degree $d_i$, is a random variable following the Poisson distribution with mean $D$. In the main text assume the threshold $\theta_i$ of node $i$ is proportional to $d_i$.  Some representative results obtained under this model assumption are shown in Figs.~4(a) and 4(b) of the main text.

\begin{figure}[h!]
  \centering
  \subfigure[]{
    \label{fig:ERk4S}
    \includegraphics[angle=270,width=0.4\textwidth]{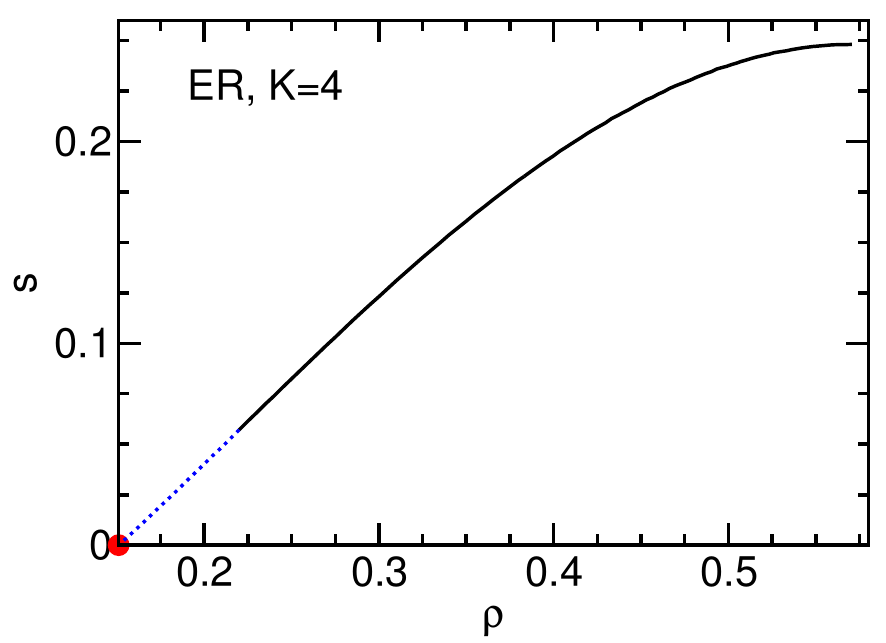}}
  \subfigure[]{
    \label{fig:ERk4SA}
    \includegraphics[angle=270,width=0.4\textwidth]{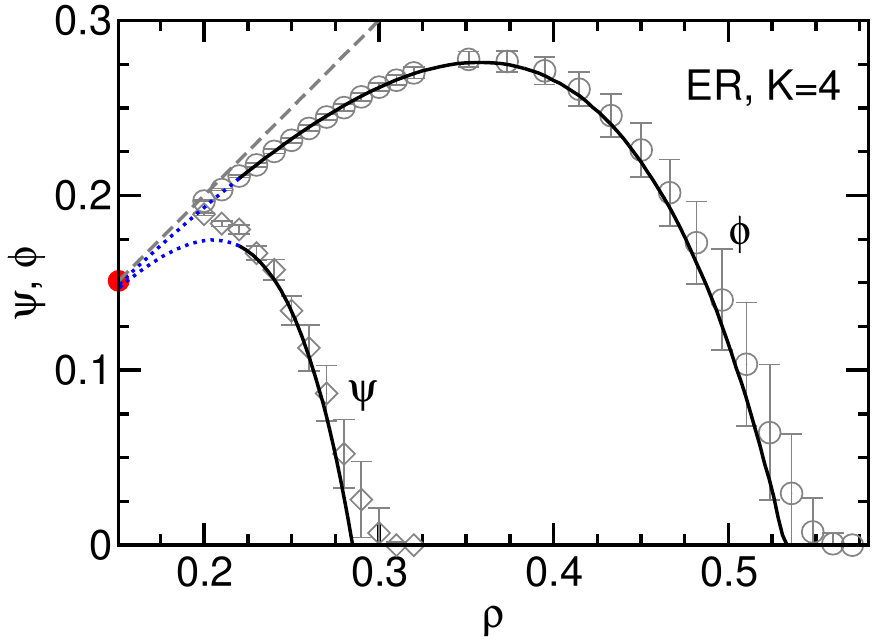}}
  \caption{
    \label{fig:ERD6k4}
    The ER random network ensemble of mean degree $D\! = \! 6$, assuming uniform threshold $K\! = \! 4$. (a): Entropy density $s$, which reaches zero at $\rho_0  = 0.1511$. (b): Order parameters $\phi$ and $\psi$. Lines are the theoretical results obtained by the mean field theory, with solid and dotted lines correspond to the concave and convex region of the entropy density, respectively. Symbols in (b) are MMC simulation results obtained on a single ER network instance of size $N \! = \! 40000$, errorbars denote standard deviations ($10^5$ alliance configurations are sampled at each energy density $\rho$). The dashed line in (b) denote the break-node fraction upper-bound $\phi = \rho$.
  }
\end{figure}

Here we report in Fig.~\ref{fig:ERD6k4} some results obtained by assuming $\theta_i$ to be uniform ($\theta_i = K$).  These results are qualitatively very similar to the results of Figs.~2(a) and 2(b). 

\subsection{Other network instances}

\begin{figure}[h!]
  \centering
  \subfigure[]{
    \label{fig:All4cores:a}
    \includegraphics[angle=270,width=0.4\textwidth]{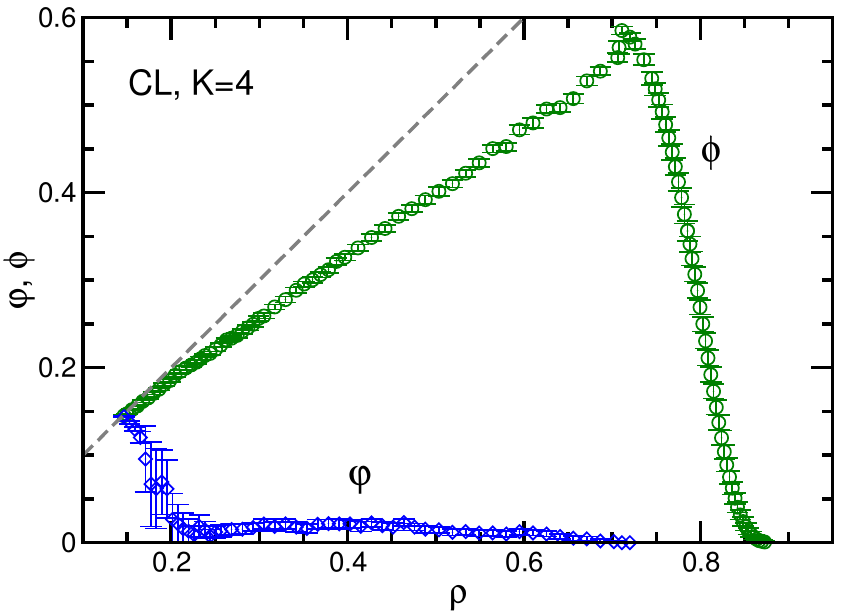}}
  \subfigure[]{
    \label{fig:All4cores:b}
    \includegraphics[angle=270,width=0.4\textwidth]{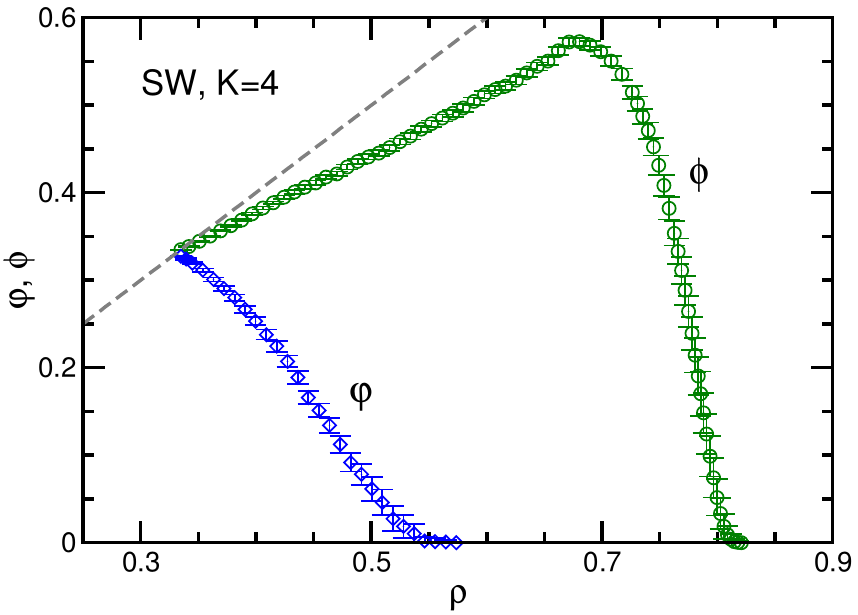}}
    
  \subfigure[]{
    \label{fig:All4cores:d}
    \includegraphics[angle=270,width=0.4\textwidth]{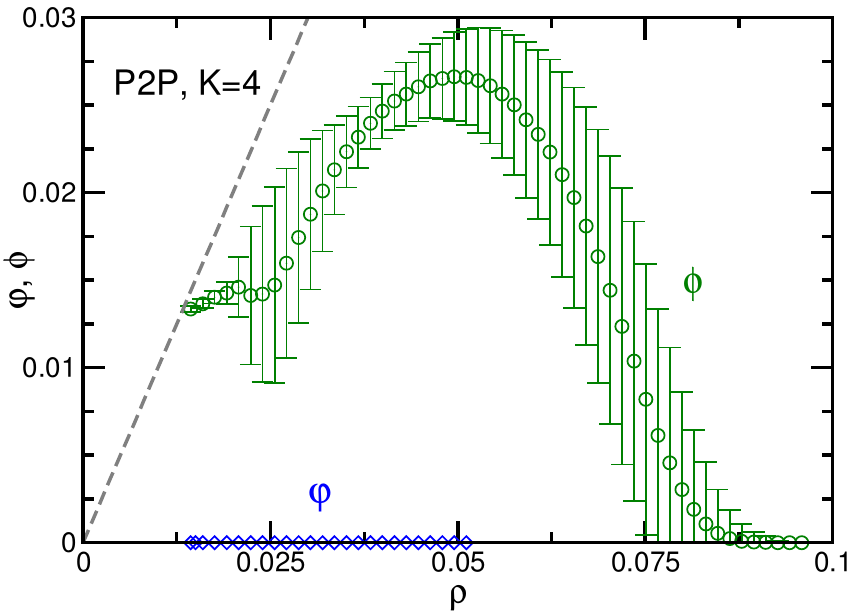}}
  \subfigure[]{
     \label{fig:All4cores:e}
    \includegraphics[angle=270,width=0.4\textwidth]{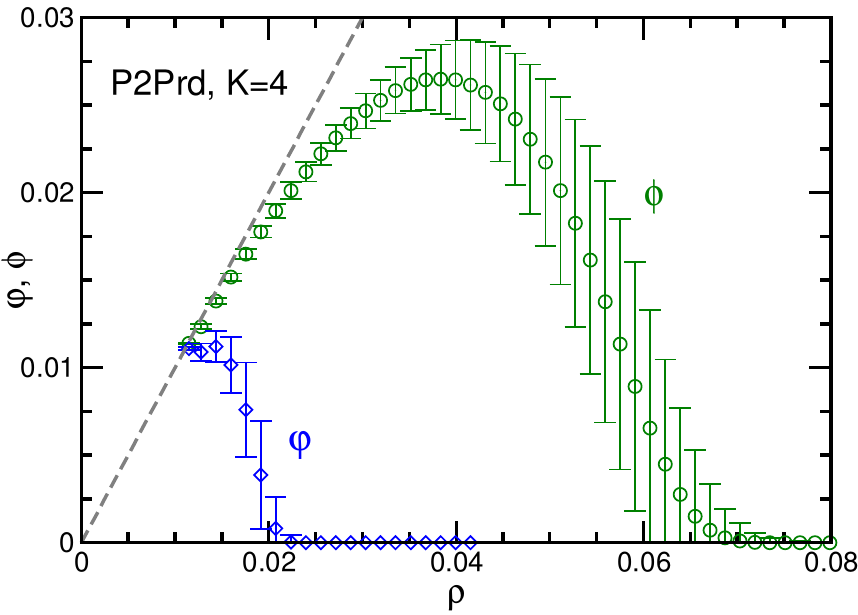}}
  \caption{
    \label{fig:All4cores}
    The fractions ($\phi$ and $\psi$) of break nodes versus energy density $\rho$ for structured networks (uniform threshold $K\! = \! 4$). (a) The periodic cubic lattice (CL) of size $N\! =\! L^3\! =\! 32768$ (side length $L\!=\! 32$ and uniform degree $D\! = \! 6$). (b) A small-world (SW) network obtained from the CL lattice by randomly rewiring $1/6$ of the links~\cite{Watts-Strogatz-1998}. (c) A peer-to-peer computer server network (P2P) with $62586$ nodes and $147892$ links, whose maximum alliance has $19765$ nodes~\cite{Ripeanu-etal-2002}. (d) The randomized P2P network (P2Prd) in which each node has the same degree as in the original P2P network but the links are completely randomized. The maximum alliance of this P2Prd network has $18747$ nodes. The symbols are MMC simulation results.  Dashed lines mark the upper bound $\phi\! = \! \rho$.
  }
\end{figure}

We show some additional simulation results obtained on random and structured networks, concerning the fractions ($\phi$ and $\psi$) of break nodes in the damage cascading processes without ($\phi$) or with ($\psi$) local recruitments (Fig.~\ref{fig:All4cores}). The networks include a periodic cubic lattice (CL, three-dimensional), a small-world (SW) network obtained from the cubic lattice by randomly rewiring $\frac{1}{6}$ of the links, and a real-world peer-to-peer (P2P) network and its link-randomized version~\cite{Ripeanu-etal-2002}, assuming fixed threshold $K\! = \! 4$. The results shown in Fig.~\ref{fig:All4cores} are similar to Figs.~2(b) and 2(c) of the main text and Fig.~\ref{fig:ERk4SA}.

Compared with the results obtained on RR and ER random networks (see Figs.~2(b) and 2(c) and Fig.~\ref{fig:ERk4SA}, respectively), we also see that the results obtained on the cubic lattice (Fig.~\ref{fig:All4cores:a}) and the small world network (Fig.~\ref{fig:All4cores:b}) have some distinctive features, namely the $\phi(\rho)$ curves seem to have a cusp close to the maximum value of $\phi$. Such a cusp may indicate some phase separation behavior. It is interesting to notice that the results on the original P2P network (Fig.~\ref{fig:All4cores:d}) and the link-reshuffled random version (Fig.~\ref{fig:All4cores:e}) are quite similar to each other, indicating that structural correlations in the original P2P network do not affect its fragility property significantly.

We also perform MMC simulations on the scientific cooperation network of condensed-matter physicists (caCondMat~\cite{Leskovec-Kleinberg-Faloutsos-2007}) and its randomized version in which all the links of the original network are reshuffled. The degree profile of the caCondMat network (and the randomized version) are quite broad (i.e., it follows a power-law approximately~\cite{Leskovec-Kleinberg-Faloutsos-2007}, referred to as scale-free~\cite{Albert-Barabasi-2002}) and many nodes are highly connected. The threshold is assumed to be the same ($K\! = \! 4$) for all the nodes. The number of nodes is $N\! =\! 23133$ and the number of links is $M\! = \! 93439$. The maximum alliance of the caCondMat network contains $13464$ nodes, while that of the randomized network contains $13564$ nodes. We find that the sampled alliance configurations for these two networks are robust to single-node perturbations ($\phi \! = \! 0$) until the relative size $\rho$ of the alliances is reaching the smallest value achievable by the MMC algorithm. To understand this strong robustness, we find that the highly connected nodes in these two networks have high probabilities of being active in a sampled alliance configuration (so that the low-degree nodes then have more freedom to be active or inactive, leading to higher entropy). These active hub nodes have a strong stabilizing effect to the alliance configurations~\cite{Dorogovtsev-etal-2006}.  We also consider synthetic scale-free random networks generated by the static method~\cite{Goh-Kahng-Kim-2001}. The degree distribution of such a purely random network decays with degree $d$ as a power-law $d^{-\gamma}$ with decay exponent $\gamma$ (we fix $\gamma \! = \! 3$ in our numerical experiments). Similar robustness behavior is observed on such random scale-free networks under the model assumption of uniform threshold $\theta_i = K$.

However, for such highly heterogeneous networks, if a node $i$ has a lot of nearest neighbors (its degree $d_i$ being large) its threshold $\theta_i$ may also be large. A plausible assumption may be to assume $\theta_i$ being proportional to $d_i$, that is,
\begin{equation}
  \label{eq:propm}
  \theta_i = r d_i
  \; ,
\end{equation}
with the ratio $r$ being a constant. This means that a node $i$ will be active only if at least a fraction $r$ of its nearest neighbors are also active. The representative results shown in Figs.~4(d) of the main text, obtained under this model assumption (\ref{eq:propm}), confirm that scale-free networks may also possess two separate tipping points $\rho_{\rm wt}$ and $\rho_{\rm st}$ as the other network types.

\end{document}